\documentclass[12pt,a4paper,english]{article}
\usepackage[T1]{fontenc}
\usepackage[latin2]{inputenc}
\usepackage[english]{babel}
\usepackage{amsmath}
\usepackage{amsfonts}
\usepackage{indentfirst}
\usepackage[dvips]{graphicx}

\newcommand{\bea}{\begin{eqnarray}}
\newcommand{\eea}{\end{eqnarray}}
\newcommand{\be}{\begin{equation}}
\newcommand{\ee}{\end{equation}}

\def\g{\gamma} 
\def\G{\Gamma}

\newcommand{\cC}{{\cal C }}

\newcommand{\cH}{{\cal H }}

\newcommand{\cZ}{{\cal Z }}

\begin{document}

\sloppy


\begin{flushright}
\begin{tabular}{l}
CALT 68-2756 \\

\\ [.3in]
\end{tabular}
\end{flushright}

\begin{center}
\Large{ \bf Wall-crossing, free fermions and crystal melting}
\end{center}

\begin{center}

\bigskip

Piotr Su{\l}kowski\footnote{On leave from University of Amsterdam and So{\l}tan Institute for Nuclear Studies, Poland.}

\bigskip

\medskip

\emph{California Institute of Technology, Pasadena, CA 91125, USA} \\ [1mm]

\emph{and} \\ [1mm]

\emph{Jefferson Physical Laboratory, Harvard University, Cambridge, MA 02138, USA} \\

\bigskip


\bigskip

\smallskip
 \vskip .4in \centerline{\bf Abstract}
\smallskip

\end{center}

We describe wall-crossing for local, toric Calabi-Yau manifolds without compact four-cycles, in terms of free fermions, vertex operators, and crystal melting. Firstly, to each such manifold we associate two states in the free fermion Hilbert space. The overlap of these states reproduces the BPS partition function corresponding to the non-commutative Donaldson-Thomas invariants, given by the modulus square of the topological string partition function. Secondly, we introduce the \emph{wall-crossing} operators which represent crossing the walls of marginal stability associated to changes of the $B$-field through each two-cycle in the manifold. BPS partition functions in non-trivial chambers are given by the expectation values of these operators. Thirdly, we discuss crystal interpretation of such correlators for this whole class of manifolds. We describe evolution of these crystals upon a change of the moduli, and find crystal interpretation of the flop transition and the DT/PT transition. The crystals which we find generalize and unify various other Calabi-Yau crystal models which appeared in literature in recent years.


\newpage 

\tableofcontents

\newpage 


\section{Introduction}   

Counting of BPS states is an important problem in supersymmetric theories. In the context of string compactifications one is interested in the spectrum of bound states of D-branes wrapped around cycles of the internal Calabi-Yau threefold. In recent years there has been much progress in understanding degeneracies of D0 and D2-branes bound to a single D6-brane in type IIA theory. It was found out that these degeneracies are related on one hand to the topological string theory, and on the other to statistical models of crystal melting \cite{ok-re-va,foam,ps-phd}, while the Donaldson-Thomas theory provides a mathematical framework to describe these developments \cite{MNOP-I}. 

In fact the number of such bound states depends on the moduli of the the underlying Calabi-Yau manifold. Once these moduli are varied certain states may either get bound together or unbound and their numbers jump. The corresponding moduli space is therefore divided into distinct chambers by loci called walls of marginal stability. The term \emph{wall-crossing} refers to understanding behavior of BPS counting functions upon crossing those walls of marginal stability. While the issue of stability in various supersymmetric theories has a long history, one of the sources of its new impetus in the context of D-branes in string theory was the work of Denef and Moore \cite{DenefMoore}. In a parallel development a very general mathematical theory which describes these phenomena has been formulated by Kontsevich and Soibelman \cite{KS}. Subsequently these results were applied to local Calabi-Yau manifolds from physical \cite{JaffMoore,ChuangJaff,RefMotQ} and mathematical \cite{Szendroi,Young,DTPT,YoungBryan,MozRei,NagaoNakajima,Nagao-strip} points of view. Recently, based on various string dualities \cite{DVV-Mduality}, these results and their relation to topological string theory were explained from the M-theory perspective for manifolds without compact four-cycles \cite{WallM}. 


In this paper we reformulate BPS counting and wall-crossing for the entire class of local, toric Calabi-Yau manifolds without compact four-cycles, in terms of free fermions, vertex operators and crystal melting.\footnote{While this paper was being prepared for publication, the author was informed that Kentaro Nagao independently reformulated wall-crossing for local Calabi-Yau manifolds in terms of vertex operators and obtained results overlapping with ours \cite{NagaoVO}. Our papers appear simultaneously.} There are several motivations for our work. One motivation is related to the wave-function interpretation of topological string theory \cite{witten-bckg,top-wave}. Such interpretation was originally proposed in the context of holomorphic anomaly equations. Various explicit representations of topological string partition functions, as states in the free fermion Hilbert space, were found in \cite{adkmv,Nek-Ok,dhsv,dhs}. However those states were constructed only in one point in the moduli space, which corresponds to the large radius limit. In \cite{adkmv} those states were related to integrable hierarchies. In \cite{dhsv} by a chain of string dualities their physical origin was found in terms of open strings stretching between intersecting D4 and D6-branes. It is therefore interesting whether some analogous wave-function interpretation extends also to other chambers of the moduli space. In this paper we show that this is indeed true. Firstly, we find a new set of fermionic states which encode the quantum structure of local, toric Calabi-Yau manifolds without compact four-cycles, such that their overlaps are equal to the modulus square of the corresponding topological string partition functions. These states are different than those found in \cite{adkmv} for the same manifolds. In particular 
the states which we find here are naturally associated to the non-commutative Donaldson-Thomas chamber. Nonetheless, both of them are given by a Bogoliubov transformation of the fermionic vacuum. Secondly, we also find fermionic interpretation of wall-crossing in a large set of chambers, and realize BPS generating functions in those chambers as various fermionic correlators. While we do not provide a clear physical explanation why such fermionic quantum states should occur, we believe it should exist.

Another motivation of our work is related to Calabi-Yau crystals. Similarly as in the wave-function case, originally a crystal interpretation of BPS counting was realized in one particular chamber of the moduli space and was intimately related to the vertex operators \cite{ok-re-va,foam}. More recently, a generating function of pyramid partitions for the resolved conifold in another special chamber was computed using free fermion formalism \cite{YoungBryan}. It is natural to expect that such fermionic crystal representation of BPS counting should hold for all chambers and much larger class of manifolds. In this paper we indeed find such representation for local, toric Calabi-Yau manifolds without compact four-cycles, in a large set of chambers.

The main results of this paper are briefly summarized below.


\subsection{Summary of the results}    \label{ssec-SummaryResults}

Let $M$ denote a toric Calabi-Yau manifolds without compact four-cycles. Firstly, we associate to $M$ two states 
$$
|\Omega_{\pm}\rangle  \in\cH
$$ 
in the Hilbert space of free fermion $\cH$, such that 
\be
\cZ \equiv |\cZ_{top}|^2 = \langle \Omega_+ | \Omega_- \rangle,
\ee
where $\cZ_{top}$ denotes the instanton part of the topological string partition function of $M$, and $\cZ$ is the generating function of the non-commutative Donaldson-Thomas invariants. 

Secondly, we find a large set of \emph{wall-crossing} operators $\overline{W}_p,\overline{W}_p'$, which inserted $n$ times in the above correlator encode the BPS generating functions in chambers corresponding to turning on $n$, respectively positive and negative, quanta of the $B$-field through $p$'th two-cycle of $M$. These generating functions, according to the prescription of \cite{WallM} are given by a restriction of $|\cZ_{top}|^2$. In our formalism, in chambers connected by a finite number of walls of marginal stability to the non-commutative Donaldson-Thomas region, or to the core region with $\widetilde{\cZ}=1$ (i.e. corresponding to positive or negative radius $R$ in the M-theory interpretation), they are given respectively by
\be
\cZ_{n|p} = \langle \Omega_+ | (\overline{W}_p)^n |\Omega_- \rangle, \qquad  \qquad \cZ_{n|p}' = \langle \Omega_+ | (\overline{W}'_p)^n |\Omega_- \rangle,
\ee
\be  
\widetilde{\cZ}_{n|p}  = \langle 0 | (\overline{W}_p)^n | 0 \rangle, \qquad \qquad \widetilde{\cZ}'_{n|p}  = \langle 0 | (\overline{W}'_p)^n | 0 \rangle.
\ee
From our point of view the generating function in the core region, corresponding to a single D6-brane, is given by
$$
\widetilde{\cZ}  =  \langle 0 | 0 \rangle = 1.
$$
In particular, the change from positive to negative values of $R$, which corresponds for example to the so-called DT/PT transition \cite{MNOP-I,DTPT}, is represented in our formalism by the change of the ground state representing the manifold
$$
|\Omega_{\pm} \rangle \quad \longleftrightarrow \quad  |0 \rangle.
$$
More precisely, the equality of the above correlators to the BPS partition functions arises upon an appropriate identification of parameters (i.e. colors of the crystal with K{\"a}hler parameters and string coupling), which we find in each case that we consider. 

Thirdly, we find a crystal melting interpretation of all these generating functions. They turn out to be related, respectively, to the generating functions 
$$
Z, \quad Z_{n|p}, \quad \widetilde{Z}_{n|p}, \quad Z_{n|p}', \quad \widetilde{Z}_{n|p}', \quad \widetilde{Z},
$$
of multi-colored crystals, also under appropriate identification of crystal and stringy parameters. The shape and coloring of these crystals is encoded in the toric diagram of the Calabi-Yau manifold, and for non-trivial chambers also in the structure of the wall operators $\overline{W}_p,\overline{W}'_p$. We also discuss evolution of crystals and find a crystal interpretation of various geometric transitions, such as the flop transition and DT/PT transition. In particular, in our framework we can easily prove the relation between certain BPS generating functions for the conifold and finite pyramid partitions conjectured in \cite{ChuangJaff}, and generalize it to other manifolds. Moreover, the crystals which we find provide a unifying point of view on various crystal models considered in literature in recent years.


\bigskip

This paper is organized as follows. In section \ref{sec-preliminaries} we review necessary background on wall-crossing, free fermion formalism and crystal melting, and set up various conventions and notation. In section \ref{sec-results} we describe in detail the results summarized above for the class of manifolds encoded in an arbitrary triangulation of a strip. In section \ref{sec-examples} we illustrate these results in several examples and find more general wall-crossing operators. We also analyze the closed topological vertex geometry, which is a case that does not arise from a triangulation of a strip. Section \ref{sec-proofs} contains proofs of the statements given in section \ref{sec-results}. Section \ref{sec-summary} contains summary and discussion.


\section{Preliminaries}    \label{sec-preliminaries}

In this section we review some background material, as well as set the notation and conventions used in further parts of the paper. In section \ref{ssec-wallcross} we review a physical picture of wall-crossing for manifolds without compact four-cycles, on which we will rely in derivation of some of our results. In section \ref{ssec-fermion} we review the formalism of free fermions and the construction of vertex operators. In section \ref{ssec-crystal} we recall how these can be used to solve certain models of three-dimensional melting crystals, which in particular arise in connection with enumerative invariants of Calabi-Yau threefolds. In section \ref{ssec-ReviewConifold} we review the wall-crossing for the resolved conifold, which we will generalize to a large class of manifolds in section \ref{sec-results}.

\subsection{Wall-crossing for local Calabi-Yau manifolds}  \label{ssec-wallcross}

In this section we review the wall-crossing phenomena for local toric Calabi-Yau manifolds without four-cycles. One large class of such manifolds can be encoded in toric diagrams which arise from a triangulation of a strip, as we will explain in detail in section \ref{ssec-FockState}. Another example of such a geometry (which does not arise from a triangulation of a strip) is the closed topological vertex presented in section \ref {ssec-ClosedVertex}. 

Mathematically wall-crossing describes a change of generalized Donaldson-Thomas invariants upon crossing the walls of marginal stability. Generating functions of such generalized Donaldson-Thomas invariants for the local geometries mentioned above, in various chambers were derived by mathematicians in \cite{Szendroi,YoungBryan,NagaoNakajima,Nagao-strip}. Physically generalized Donaldson-Thomas invariants correspond to numbers of D6-D2-D0 bound states, and for the resolved conifold they were analyzed in \cite{JaffMoore,ChuangJaff,RefMotQ}. Recently a physical prescription for determining BPS generating functions in all chambers, for general manifolds without compact four-cycles, was derived in \cite{WallM}. For local manifolds mentioned above this prescription agrees with mathematical results. This also makes contact with topological string theory and fits naturally in the context of our results. We therefore review the wall-crossing for local geometries from the perspective of this paper.

The idea of \cite{WallM} is as follows. The system of D6-D2-D0 branes in type IIA theory can be lifted to M-theory on $S^1$, whereupon D6-brane transforms into a geometric background of a Taub-NUT space with unit charge, extending in directions transverse to the D6-brane. This Taub-NUT space is a circle $S^1_{TN}$ fibration over $\mathbb{R}^3$, with $S^1_{TN}$ shrinking to a point in the location of the D6-brane and attaining a radius $R$ at infinity. The counting of bound states involving D6-brane is then reinterpreted as the counting of BPS states of M2-branes in this Taub-NUT space. This counting does not change when the radius $R$ grows to infinity and Taub-NUT approaches $R^4$. Moreover, from the spacetime perspective, the resulting generating functions of BPS degeneracies would factorize into a product of single particle partition functions if there would be no interaction among five-dimensional particles. It is then argued that this can be achieved if the following two conditions are satisfied. Firstly, the moduli of the Calabi-Yau have to be tuned such that M2-branes wrapped in various ways have aligned central charges. This can be achieved by considering vanishing K{\"a}hler parameters of the Calabi-Yau space. At the same time, to avoid generation of massless states, non-trivial flux of the M-theory three-form field through the two-cycles of the Calabi-Yau and $S^1_{TN}$ have to be turned on. In type IIA this flux translates to the $B$-field flux $B$ through two-cycles of Calabi-Yau. For a state arising from $D2$ wrapping a class $\beta$ the central charge then reads
\be
Z(l,\beta) = \frac{1}{R}(l + B\cdot \beta),         \label{Zcentral}
\ee
where $l$ counts the D0-brane charge, which is taken positive to preserve the same supersymmetry. The second condition requires that the only BPS states in the Taub-NUT geometry are particle-like, and therefore there are no string-like states which would arise from M5-branes wrapped on four-cycles. This is why the results of \cite{WallM} hold for Calabi-Yau geometries without compact four-cycles. 

Under the above two assumptions, and in the limit $R\to\infty$, the counting of D6-D2-D0 bound states translates to the counting of particle degeneracies on $\mathbb{R}^4\times S^1$, arising from M2-branes wrapped on cycles $\beta$. The excitations of these particles in $\mathbb{R}^4$, parametrized by two complex variables $z_1,z_2$, are accounted for by the modes of the holomorphic field 
$$
\Phi(z_1,z_2) = \sum_{l_1,l_2} \alpha_{l_1,l_2} z_1^{l_1} z_2^{l_2}.
$$
Under a decomposition of the isometry group of $\mathbb{R}^4$ as $SO(4)=SU(2)\times SU(2)'$ there are $N_{\beta}^{m,m'}$ five-dimensional BPS states of intrinsic spin $(m,m')$. The degeneracies we are interested in correspond to the net number obtained by tracing over the $SU(2)'$ spins, and are expressed by Gopakumer-Vafa invariants
$$
N_{\beta}^m = \sum_{m'} (-1)^{m'} N_{\beta}^{m,m'}.
$$
The total angular momentum of a given state contributing to the index is $l=l_1+l_2+m$. Now the invariant degeneracies are expressed as the trace over the corresponding Fock space, subject to the condition that all the contributing states are mutually BPS, i.e.
\be
Z(l,\beta) = \frac{1}{R}(l + B\cdot \beta) > 0 .       \label{Zpositive}
\ee
Therefore, in a chamber specified by the moduli $R$ and $B$
\bea
\cZ(R,B) & = & \textrm{Tr}_{Fock} q_s^{Q_0} Q^{Q_2} |_{Z(l,\beta)>0} = \nonumber \\
& = & \prod_{\beta,m} \prod_{l_1+l_2 = l} (1-q_s^{l_1+l_2+m} Q^{\beta})^{N_{\beta}^{m}} | _{Z(l,\beta)>0} \nonumber \\
& = & \prod_{\beta,m} \prod_{l=1}^{\infty} (1-q_s^{l+m} Q^{\beta})^{l N_{\beta}^{m}} | _{Z(l,\beta)>0},     \label{cZ-chamber}
\eea
where $Q=e^{-T}$ and $q_s=e^{-g_s}$ encode respectively the K{\"a}hler class $T$ and the string coupling $g_s$. Note that the product over $\beta$ runs over both positive and negative classes, so that both M2 and anti-M2-branes contribute to the index as long as the condition (\ref{Zpositive}) is satisfied.

An important observation in \cite{WallM} is also the fact that the BPS generating functions are simply related to the topological string partition function
\be
\cZ(R,B) = \cZ_{top}(Q) \cZ_{top}(Q^{-1}) |_{chamber} \equiv |\cZ_{top}(Q)|^2|_{chamber} .     \label{ZBPS-Ztop}
\ee
Here the subscript $|_{chamber}$ denotes restriction to contributions from states for which $Z(l,\beta)>0$ in a given chamber, and the topological string partition function is expressed through Gopakumar-Vafa invariants
$$
\cZ_{top}(Q) = M(q)^{\chi/2} \prod_{l=1}^{\infty} \prod_{\beta>0, m} (1 - Q^{\beta} q_s^{m+l})^{l N^m_{\beta}},
$$
where $M(q) = \prod_l (1-q^l)^{-l}$ is the MacMahon function and $\chi$ is the Euler characteristic of the Calabi-Yau manifold.

There are a few interesting special cases of the relation (\ref{ZBPS-Ztop}). For positive $R$ and infinite $B$ we get contributions from states with arbitrary $n$ and positive $\beta$, so that we get
\be
\cZ(R>0,B\to\infty) = M(1,q)^{\chi/2} \cZ_{top}(Q).   \label{chamber-RposBinf}
\ee
This immediately leads to the relation between the Gromov-Witten and Donaldson-Thomas invariants discussed in \cite{foam}. For positive $R$ and $B$ sufficiently small $0<B<<1$ all $n$ and $\beta$ contribute, which corresponds to the chamber for which non-commutative Donaldson-Thomas invariants \cite{Szendroi} are defined
\be
\cZ(R>0,0<B<<1) = \cZ_{top}(Q) \cZ_{top}(Q^{-1}) \equiv |\cZ_{top}(Q)|^2.        \label{chamber-RposBsmall}
\ee 
For negative $R$ and $B$ sufficiently small, only a single D6-brane contributes to the partition function
\be
\widetilde{\cZ}(R<0,0<B<<1) = 1.        \label{chamber-singleD6}
\ee

In general in what follows we will denote BPS generating functions in chambers with positive $R$ by (curly) $\cZ$, and in chambers with negative $R$ by $\widetilde{\cZ}$. In this paper we will show that they are equal to certain free fermion amplitudes, as well as crystal model generating functions denoted respectively by (non-curly) $Z$ and $\widetilde{Z}$, under a simple identification of parameters on both sides.


\subsection{Free fermion formalism}  \label{ssec-fermion}

Formalism of free fermions in two dimensions is well known \cite{jimbo-miwa,macdonald} and ubiquitous in literature on topological strings and crystal melting \cite{ok-re-va,ps-phd,YoungBryan,adkmv,dhsv}. The main purpose of this section is therefore to set up a notation which we will follow in the remaining parts of this paper. Our conventions follow closely those of \cite{YoungBryan}.

The states in the free fermion Fock space are created by the action of the anti-commuting modes of the fermion field 
$$
\psi(z) = \sum_{k\in\mathbb{Z}} \psi_{k+1/2} z^{-k-1}, \qquad \psi^*(z) = \sum_{k\in\mathbb{Z}} \psi^*_{k+1/2} z^{-k-1},  \qquad \{ \psi_{k+1/2},\psi^*_{-l-1/2} \} = \delta_{k,l}
$$ 
on the vacuum $|0\rangle$. Each state 
$$
|\mu\rangle = \prod_{i=1}^d \psi^*_{-a_i-1/2} \psi_{-b_i-1/2} |0\rangle, \qquad \textrm{with} \quad  a_i = \mu_i-i,\ b_i =\mu^t_i-i,
$$
corresponds in a unique way to a two-dimensional partition $\mu=(\mu_1,\mu_2,\ldots \mu_l)$. The modes $\alpha_m$ of the bosonized field $\partial \phi = :\psi(z)\psi^*(z):$ satisfy the Heisenberg algebra $[\alpha_m,\alpha_{-n}] = n \delta_{m,n}.$

One can now define vertex operators
\be
\G_{\pm}(x) = e^{\sum_{n>0} \frac{x^n}{n}\alpha_{\pm n}}, \qquad \qquad \G'_{\pm}(x) = e^{\sum_{n>0} \frac{(-1)^{n-1}x^n}{n}\alpha_{\pm n}},     \label{vertexoperators}
\ee
which act on fermionic states $|\mu\rangle$ corresponding to partitions $\mu$ as \cite{jimbo-miwa,macdonald,YoungBryan}
\bea
\G_-(x) |\mu\rangle  =  \sum_{\lambda \succ \mu} x^{|\lambda|-|\mu|}|\lambda\rangle,  & &\qquad \qquad  
\G_+(x) |\mu\rangle  =  \sum_{\lambda \prec \mu} x^{|\mu|-|\lambda|}|\lambda\rangle,   \label{Gmu}   \\
\G'_-(x) |\mu\rangle =  \sum_{\lambda^t \succ \mu^t} x^{|\lambda|-|\mu|}|\lambda\rangle, & & \qquad \qquad
\G'_+(x) |\mu\rangle  =  \sum_{ \lambda^t \prec \mu^t} x^{|\mu|-|\lambda|}|\lambda\rangle,   \label{Gprimmu}
\eea
where the interlacing relation between partitions is defined by
\be
\lambda \succ \mu  \qquad  \Leftrightarrow \qquad  \lambda_1 \geq \mu_1 \geq \lambda_2 \geq \mu_2 \geq \lambda_3 \geq \ldots.  \label{interlace}
\ee

The operator $\G'$ is in fact the inverse of $\G$ with negative argument. These operators satisfy commutation relations
\bea
\G_+(x) \G_-(y) & = & \frac{1}{1-xy} \G_-(y) \G_+(x),   \label{GplusGminus}\\
\G'_+(x) \G'_-(y) & = & \frac{1}{1-xy} \G'_-(y) \G'_+(x),   \\
\G'_+(x) \G_-(y) & = & (1+xy) \G_-(y) \G'_+(x),   \\
\G_+(x) \G'_-(y) & = & (1+xy) \G'_-(y) \G_+(x).
\eea

We also introduce various colors $q_g$ and the corresponding operators $\widehat{Q}_g$ (a hat is to distinguish them from K{\"a}hler parameters $Q_i$)
\be
\widehat{Q}_g|\lambda\rangle = q_g^{|\lambda|}|\lambda\rangle.     \label{Qcolor}
\ee
These operators commute with vertex operators up to a scaling of the argument
\bea
\G_+(x) \widehat{Q}_g = \widehat{Q}_g \G_+(x q_g), & & \qquad \qquad
\G'_+(x) \widehat{Q}_g = \widehat{Q}_g \G'_+(x q_g), \\
\widehat{Q}_g \G_-(x) = \G_-(x q_g) \widehat{Q}_g, & & \qquad \qquad 
\widehat{Q}_g \G'_-(x) = \G'_-(x q_g) \widehat{Q}_g.    \label{Qcommute}
\eea


\subsection{Crystal melting}    \label{ssec-crystal}

Since the work \cite{ok-re-va,foam} it is known that various enumerative invariants of Calabi-Yau manifolds turn out to be related to statistical models of crystal melting. More precisely, generating functions of those enumerative invariants are equal to partition functions of crystal models in the grand canonical ensemble. Such crystal partition functions are computed as sums over all possible crystal configurations subject to appropriate rules, with weights given by the number (or its refinements) of (missing) elementary crystal constituents in a given configuration. 

Enumerative invariants of Calabi-Yau threefolds are related to three-dimensional crystal models. A simple example of such a model consists of unit cubes filling the positive octant of $\mathbb{R}^3$ space. A unit cube located in position $(I,J,K)$ can evaporate from this crystal only if all other cubes with coordinates $(i\leq I,j\leq J,k\leq K)$ already evaporated. Therefore all missing configurations are in one-to-one correspondence with three-dimensional partitions $\pi$, also called plane partitions. Weighting each such configuration by the number of boxes it consists of $|\pi|$, the partition function of this model turns out to be the MacMahon function $M(q)\equiv M(1,q)$
$$
Z = \sum_{\pi} q^{|\pi|} = \sum_{l=0}^{\infty} p(l) q^l = \prod_{l=1}^{\infty}  \frac{1}{(1-q^l)^l} = M(q).
$$
Mathematically the numbers $p(l)$ encode Donaldson-Thomas invariants, while from the string theory viewpoint they count the number of bound states of $l$ D0-branes with a single D6-brane covering $\mathbb{C}^3$ \cite{ok-re-va,foam}. In what follows we also often use generalized MacMahon functions
\be
M(x,q) = \prod_{i=1}^{\infty} (1-xq^i)^{-i}, \qquad \qquad  \widetilde{M}(x,q) = M(x,q) M(x^{-1},q).   \label{MacMahons}
\ee

\begin{figure}[htb]
\begin{center}
\includegraphics[width=\textwidth]{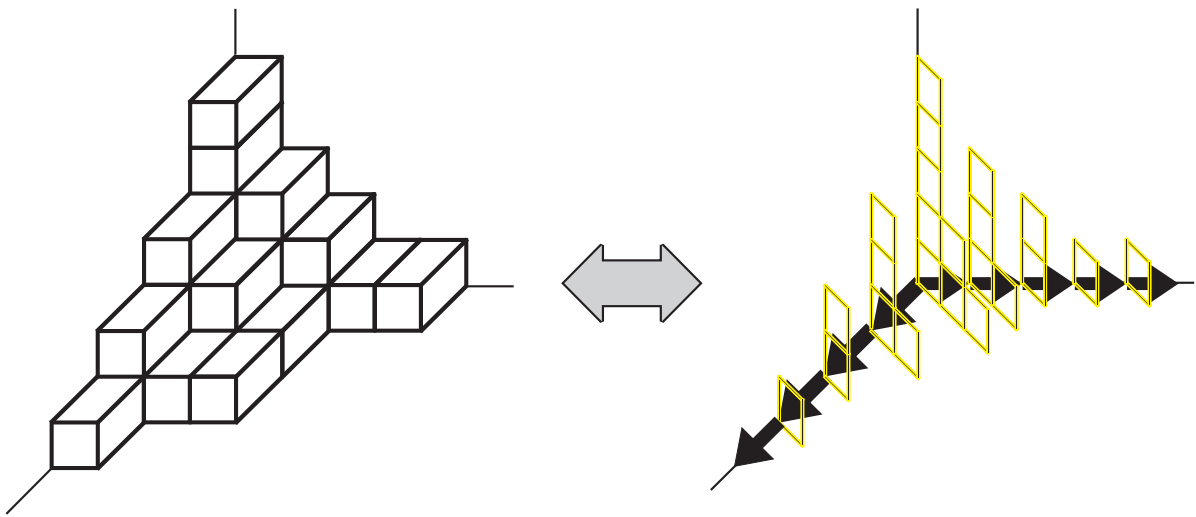} 
\begin{quote}
\caption{\emph{Slicing of a plane partition (left) into a sequence of interlacing two-dimensional partitions (right). A sequence of $\G_{\pm}$ operators in (\ref{C3-crystal}) which create two-dimensional partitions is represented by arrows inserted along two axes. Directions of arrows $\rightarrow$  represent interlacing condition $\succ$ on partitions. We reconsider this example from a new viewpoint in figure \ref{fig-C3new}.}} \label{fig-C3}
\end{quote}
\end{center}
\end{figure}

The above partition function $Z$ is a prototype example which can be computed using free fermion formalism and vertex operators. Denoting the axes of $\mathbb{R}$ in which the crystal is embedded by $x,y,z$, we first slice each possible crystal configuration (or rather its missing complement) by planes given by $x-y \in\mathbb{Z}$, as shown in figure \ref{fig-C3}. One can show that a configuration of boxes in each such slice corresponds to a two-dimensional partition, and two such partitions, corresponding to two neighboring slices, satisfy the interlacing condition (\ref{interlace}). Therefore constructing all possible crystal configurations is equivalent to building them slice by slice from interlacing partitions. Precisely such an operation is performed by $\G_{\pm}$ operators (\ref{Gmu}). Counting of boxes in a given plane partition can also be performed slice by slice using single operator $\widehat{Q}$ defined in (\ref{Qcolor}). Therefore the above partition function can be computed by writing infinite series of operators $\G_{\pm}(1) \widehat{Q}$ acting on two vacua representing empty two-dimensional diagrams, and then commuting them according to (\ref{GplusGminus}) and (\ref{Qcommute}) 
\bea
Z & = & \langle 0 | \ldots \widehat{Q} \G_+(1) \widehat{Q} \G_+(1) \widehat{Q} \G_+(1) \widehat{Q} \G_-(1) \widehat{Q} \G_-(1) \widehat{Q} \G_-(1) \widehat{Q} \ldots | 0 \rangle =  \label{C3-crystal} \\
& = & \langle 0 | \ldots \G_+(q^2) \G_+(q) \G_+(1)  \G_-(q)  \G_-(q^2) \G_-(q^3) \ldots | 0 \rangle =  \nonumber \\
& = & \prod_{l_1,l_2=1}^{\infty} \frac{1}{1-q^{l_1+l_2-1}} = M(q).   \nonumber
\eea
This computation is represented in figure \ref{fig-C3} (right), with arrows representing insertions of $\G_{\pm}$ operators along two axes. 

The example presented above is very simple, as it involves just one color $\widehat{Q}$ and only $\G_{\pm}$ operators. 
More complicated crystal models can involve more colors, as well as both $\G_{\pm}$ and $\G'_{\pm}$ operators. A family of multicolored crystals (realized by multicolored plane partitions) corresponding to $\mathbb{C}^3/\mathbb{Z}_N$ orbifolds was considered in \cite{YoungBryan}. On the other hand the crystal model encoding generalized Donaldson-Thomas invariants for the conifold involves two-colored pyramid partitions which are built by an application of interlacing sequence of $\G$ and $\G'$ operators, as we will discuss in more detail in sections \ref{ssec-ReviewConifold} below and \ref{ssec-ExampleConifold}. In section \ref{sec-results} we find a unifying viewpoint on all these examples.


\subsection{Wall-crossing for the conifold and pyramid partitions}   \label{ssec-ReviewConifold}

In this section we briefly review the wall-crossing for the resolved conifold. The structure of walls and chambers has been analyzed in this case in \cite{JaffMoore,ChuangJaff,Szendroi,NagaoNakajima}, and it is of course consistent with the results in \cite{WallM} summarized in section \ref{ssec-wallcross}. For each chamber there is also an associated crystal model of two-colored pyramid partitions, and crossing a wall of marginal stability corresponds to the extension of the pyramid crystal. One can also translate counting of pyramid partitions into a dimer model; in this language crossing a wall corresponds to a combinatorial operation called \emph{dimer shuffling}. The structure of these pyramid crystal models or dimers can also be encoded in a quiver and an associated potential. 

\begin{figure}[htb]
\begin{center}
\includegraphics[width=0.9\textwidth]{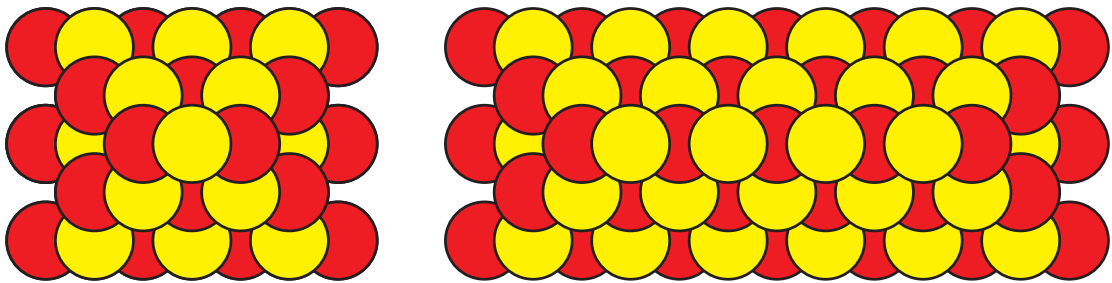} 
\begin{quote}
\caption{\emph{Infinite pyramids with one (left) and four (right) stones in the top row. Their generating functions are given respectively by $Z_0^{pyramid}$ and $Z_3^{pyramid}$.}} \label{fig-pyramids-coni-inf}
\end{quote}
\end{center}
\end{figure}

The space of stability conditions of the conifold can be divided into several infinite countable sets of chambers. As explained in section \ref{ssec-wallcross}, BPS generating functions in all chambers can also be related to the (square of the) topological string partition function
\be
\cZ^{conifold}_{top} (Q) = M(1,q_s)  \prod_{k\geq 1}(1 - Q q_s^k)^k   \label{Ztop-conifold}
\ee
with appropriate values of the moduli $R$ and the $B$ field through $\mathbb{P}^1$ of the conifold. Here $q_s=e^{-g_s}$ and K{\"a}hler modulus $T$ is encoded in $Q=e^{-T}$. This topological string partition function encodes the Gopakumar-Vafa invariants
$$
N^0_{\beta=0} = -2, \qquad \qquad N^0_{\beta = \pm 1} = 1.
$$

For future reference we write down BPS generating functions for two sets of chambers below. We stress that the labeling of the chambers, as well as an identification of the pyramid colors $q_0,q_1$ with string parameters $q_s,Q$ is slightly different than in \cite{JaffMoore,NagaoNakajima}. The conventions we use are well motivated from the point of view of the formalism developed in section \ref{sec-results}. 

The first set of chambers is characterized by $R>0$ and positive $B\in ]n,n+1 [$ (for $n\geq 0$). It extends between the non-commutative region of Szendroi (\ref{chamber-RposBsmall}) and the chamber with standard Donaldson-Thomas invariants (\ref{chamber-RposBinf}). BPS partition functions are labeled by $n$ and read
\be
\cZ_{n}^{conifold} = M(1,q)^2 \prod_{k\geq 1}(1 - Q q_s^k)^k  \prod_{k\geq n+1}(1 - Q^{-1} q_s^k)^{k} .   \label{ZnDT}
\ee
These partition functions are related to pyramid partitions with two colors $q_0$ and $q_1$, presented in figure \ref{fig-pyramids-coni-inf}. The generating function of a pyramid with $n+1$ yellow boxes in its top row is
\be
Z_n^{pyramid}(q_0,q_1) = M(1,q_0 q_1)^2 \prod_{k\geq n+1}(1 + q_0^k q_1^{k+1})^{k-n} \prod_{k\geq 1}(1 + q_0^k q_1^{k-1})^{k+n}   \label{Zn}
\ee
and it reproduces $\cZ_{n}^{conifold} \equiv Z_n^{pyramid}$ under the identification of parameters
$$
\cZ_n^{conifold} \ \textrm{chambers}:  \quad q_s = q_0 q_1, \quad  Q = -q_s^n q_1.
$$
For $n=0$ the non-commutative Donaldson-Thomas partition function \cite{Szendroi,Young} corresponds to a pyramid with just one stone in the top row, while $n\to \infty$ corresponds to the pyramid which looks like a half-infinite prism. 

\begin{figure}[htb]
\begin{center}
\includegraphics[width=0.6\textwidth]{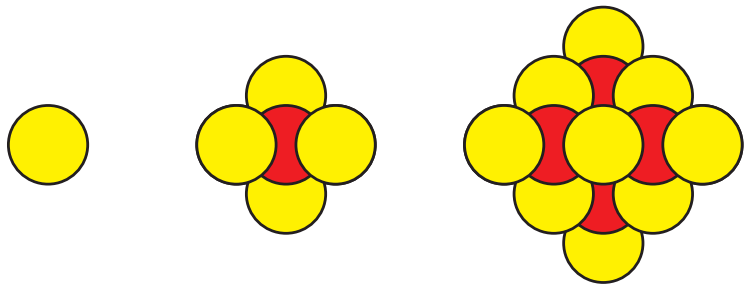} 
\begin{quote}
\caption{\emph{Finite pyramids with $m=1,2,3$ stones in the top row (respectively left, middle and right), whose generating functions are given by $\widetilde{Z}_{m+1}^{pyramid}$ (note that $\widetilde{Z}_1^{pyramid}=1$ corresponds to an empty pyramid corresponding to the pure D6-brane).}} \label{fig-pyramids-coni-fin}
\end{quote}
\end{center}
\end{figure}

The second set of chambers is characterized by $R<0$ and positive $B\in ]n-1,n [$ (for $n\geq 1$). It extends between the core region with a single D6-brane (\ref{chamber-singleD6}) and the chamber characterized by so-called Pandharipande-Thomas invariants (for the flopped geometry, or equivalently for anti-M2-branes); the BPS generating functions read
\be
\widetilde{\cZ}_{n}^{conifold}  = \prod_{j=1}^{n-1} \big(1 - \frac{q_s^j}{Q} \big)^j.  \label{ZZnDT}
\ee
The corresponding statistical models were conjectured in \cite{ChuangJaff} to correspond to finite pyramids with $n-1$ stones in the top row, as shown in figure \ref{fig-pyramids-coni-fin}. In section \ref{ssec-ExampleConifold} we will provide a new proof of this statement,\footnote{This statement has been proved also by mathematicians in \cite{NagaoNakajima}.} and show that the generating functions of such partitions are equal to
\be
\widetilde{Z}_{n}^{pyramid}(q_0,q_1) = \prod_{j=1}^{n-1} (1 + q_0^{n-j} q_1^{n-j-1} )^j.  \label{ZZn}
\ee
The equality $\widetilde{\cZ}_{n}^{conifold}  \equiv \widetilde{Z}_n^{pyramid} $ arises upon an identification
$$
\widetilde{\cZ}_n^{conifold}  \ \textrm{chambers}: \quad q_s^{-1} = q_0 q_1, \quad Q = -q_s^{n} q_1.
$$

There are two more sets of chambers characterized by the negative value of the $B$-field. The corresponding generating functions are of the analogous form as above, but with $Q$ replaced by $Q^{-1}$. We will see that in a natural way they correspond to pyramids with a \emph{vertical} top row consisting of \emph{red} stones (rather than the horizontal yellow top rows as in figures \ref{fig-pyramids-coni-inf} and \ref{fig-pyramids-coni-fin}, even though such a correspondence would also be possible, albeit with less natural identification of parameters).


\section{Results}  \label{sec-results}

In this section we consider wall-crossing for local, toric Calabi-Yau manifolds without compact four-cycles. We reformulate it in the free fermion framework and find the corresponding crystal melting picture, as anticipated in section \ref{ssec-SummaryResults}. Unless otherwise stated, we use notation and conventions introduced in section \ref{sec-preliminaries}.

There are two classes of local, toric Calabi-Yau manifolds without compact four-cycles. The first class corresponds to manifolds whose toric diagram is given by the so-called triangulation of a strip. The second class consists just of the closed topological vertex and its flop, which do not arise from a triangulation of a strip. Nonetheless the case of the closed topological vertex is closely related to one particular manifold from the first class. For this reason we focus now mainly on the first class of manifolds, and discuss the closed topological vertex later on in section \ref{ssec-ClosedVertex}.


\subsection{Triangulations and associated operators}         \label{ssec-triangulate}

To reformulate wall-crossing in the fermionic language we associate first several operators to a given local, toric manifold which arises from a triangulation of a strip.

We recall first that such manifolds arise from a triangulation, into triangles of area $1/2$, of a long rectangle or a strip of height 1. A toric diagram arises as a dual graph to such a triangulation. Each $\mathbb{P}^1$ in a Calabi-Yau geometry, represented by a finite interval in a toric diagram, corresponds to an inner line in a strip triangulation. From each vertex in a toric diagram emanates one semi-infinite vertical line, which crosses either the upper or the lower edge of the strip. Two such consecutive lines can emanate either in the same or in the opposite direction, respectively when they are the endpoints of an interval representing $\mathbb{P}^1$ with local $\mathcal{O}(-2)\oplus\mathcal{O}$ or $\mathcal{O}(-1)\oplus\mathcal{O}(-1)$ neighborhood. 

We introduce the following notation. We number independent $\mathbb{P}^1$'s from 1 to $N$, starting from the left end of the strip, and denote their K{\"a}hler parameters by $Q_i=e^{-T_i}$, $i=1,\ldots,N$. We also number, starting from the left, all vertices in a toric diagram, and associate to each vertex its type $t_i=\pm 1$, in the following way\footnote{In the same way the type A or B was associated to vertices in a triangulation of a strip in \cite{strip}.}: if the local neighborhood of $\mathbb{P}^1$, represented by an interval between vertices $i$ and $i+1$, is $\mathcal{O}(-2)\oplus\mathcal{O}$, then $t_{i+1}=t_i$; if this neighborhood is of $\mathcal{O}(-1)\oplus\mathcal{O}(-1)$ type, then $t_{i+1}=-t_i$. The type of the first vertex could be chosen arbitrarily, but to fix attention we set $t_1=+1$. We also recall that the instanton part of the closed topological string partition function for such geometries reads \cite{strip}
\be
\cZ_{top}(Q_i) =  M(1,q)^{\frac{N+1}{2}} \prod_{l=1}^{\infty} \prod_{1\leq i < j \leq N+1} \Big(1- q^l\, (Q_i Q_{i+1}\cdots Q_{j-1}) \Big)^{-(t_i t_j) l}.   \label{Ztop-strip}
\ee

\begin{figure}[htb]
\begin{center}
\includegraphics[width=0.9\textwidth]{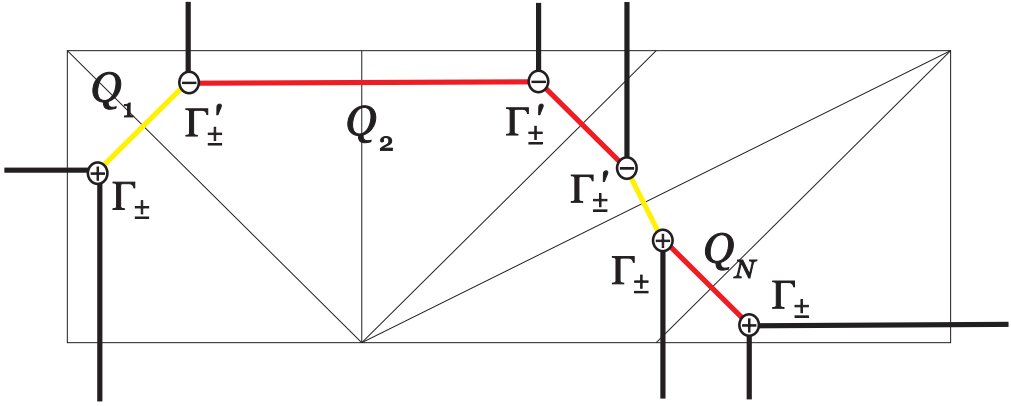} 
\begin{quote}
\caption{\emph{Toric Calabi-Yau manifolds represented by a triangulation of a strip. There are $N$ independent $\mathbb{P}^1$'s with K{\"a}hler parameters $Q_i=e^{-T_i}$, and $N+1$ vertices to which we associate $\G$ and $\G'$ operators represented respectively by $\oplus$ and $\ominus$ signs. Yellow intervals, which connect vertices with opposite signs, represent $\mathcal{O}(-1)\oplus\mathcal{O}(-1)\to \mathbb{P}^1$ local neighborhoods. Red intervals, which connect vertices with the same signs, represent $\mathcal{O}(-2)\oplus\mathcal{O}\to \mathbb{P}^1$ local neighborhoods. The first vertex on the left is chosen to be $\oplus$.}} \label{fig-strip}
\end{quote}
\end{center}
\end{figure}

Below we introduce several operators which are the main building blocks of the fermionic and crystal construction. Their structure is encoded in the toric diagram of the manifold described above. These operators are given by a string of $N+1$ vertex operators $\G_{\pm}^{t_i}(x)$ (introduced in (\ref{vertexoperators})) which we associate to the vertices of the toric diagram, and determine their type by the type of these vertices $t_i$ such that
$$
\G_{\pm}^{t_i=+1}(x)=\G_{\pm}(x),\qquad \qquad \G_{\pm}^{t_i=-1}(x)=\G_{\pm}'(x).
$$ 
Moreover this string of $\G_{\pm}^{t_i}(x)$ operators is interlaced with $N+1$ operators $\widehat{Q}_i$ representing colors $q_i$, for $i=0,1,\ldots,N$. Operators $\widehat{Q}_1,\ldots,\widehat{Q}_N$ are associated to $\mathbb{P}^1$ in the toric diagram, and there is an additional $\widehat{Q}_0$. We also define 
\be
\widehat{Q} = \widehat{Q}_0 \widehat{Q}_1\cdots \widehat{Q}_N,\qquad \qquad q = q_0 q_1 \cdots q_N.
\ee
To sum up, the upper indices of $\G_{\pm}^{t_i}(x)$ and a choice of colors of the operators which we introduce below are specified by the data of the toric manifold under consideration. The lower indices $\pm$ will be related to the wall-crossing chamber we will be interested in.

Now we can introduce the operators of interest. The first two are defined as
\be
\overline{A}_{\pm}(x) = \G_{\pm}^{t_1} (x) \widehat{Q}_1 \G_{\pm}^{t_2} (x) \widehat{Q}_2 \cdots \G_{\pm}^{t_N} (x) \widehat{Q}_N \G_{\pm}^{t_{N+1}} (x) \widehat{Q}_0.                  \label{Apm}
\ee
Commuting all $\widehat{Q}_i$'s to the left or right using (\ref{Qcommute}) we also introduce related operators
\bea
A_+(x) & = & \widehat{Q}^{-1} \, \overline{A}_{+}(x) = \G_{+}^{t_1} \big(xq\big)  \G_{+}^{t_2} \big(\frac{xq}{q_1}\big) \G_{+}^{t_3} \big(\frac{xq}{q_1 q_2}\big) \cdots \G_{+}^{t_{N+1}} \big(\frac{xq}{q_1q_2\cdots q_N}\big), \label{Aplus} \\
A_-(x) & = & \overline{A}_{-}(x) \, \widehat{Q}^{-1} = \G_{-}^{t_1} (x)  \G_{-}^{t_2} (xq_1) \G_{-}^{t_3} (x q_1 q_2) \cdots \G_{-}^{t_{N+1}} (x q_1q_2 q_N). \label{Aminus}
\eea

Secondly, we define operators which we will refer to as the \emph{wall-crossing operators}
\begin{align}
& \overline{W}_p(x) =  \Big(\G_{-}^{t_1} (x) \widehat{Q}_1 \G_{-}^{t_2} (x) \widehat{Q}_2 \cdots \G_{-}^{t_p} (x) \widehat{Q}_p\Big)\Big( \G_{+}^{t_{p+1}} (x) \widehat{Q}_{p+1} \cdots \G_{+}^{t_N} (x) \widehat{Q}_N \G_{+}^{t_{N+1}} (x) \widehat{Q}_0 \Big)  \label{Wp} \\
& \overline{W}_p'(x)  =  \Big(\G_{+}^{t_1} (x) \widehat{Q}_1 \G_{+}^{t_2} (x) \widehat{Q}_2 \cdots \G_{+}^{t_p} (x) \widehat{Q}_p\Big)\Big( \G_{-}^{t_{p+1}} (x) \widehat{Q}_{p+1} \cdots \G_{-}^{t_N} (x) \widehat{Q}_N \G_{-}^{t_{N+1}} (x) \widehat{Q}_0 \Big)   \label{Wpprim}  
\end{align}
The order of $\G$ and $\G'$ is the same as for $\overline{A}_{\pm}$ operators; the only difference is that now there are subscripts $\mp$ on first $p$ operators and $\pm$ on the remaining ones. 

In what follows we also use the following auxiliary operators, defined for fixed $p$ (corresponding to fixed $p$'th $\mathbb{P}^1$ in the toric geometry):
\bea
\g^1_+(x) & = & \G_{+}^{t_1} \big(xq\big)  \G_{+}^{t_2} \big(\frac{xq}{q_1}\big) \G_{+}^{t_3} \big(\frac{xq}{q_1 q_2}\big) \cdots \G_{+}^{t_{p}} \big(\frac{xq}{q_1q_2\cdots q_{p-1}}\big), \nonumber \\
\g^2_+(x) & = & \G_{+}^{t_{p+1}} \big(\frac{xq}{q_1 q_2\cdots q_p}\big) \cdots \G_{+}^{t_{N+1}} \big(\frac{xq}{q_1q_2\cdots q_{N}}\big), \nonumber \\
\g^1_-(x) & = & \G_{-}^{t_1} (x)  \G_{-}^{t_2} (xq_1) \G_{-}^{t_3} (x q_1 q_2) \cdots \G_{-}^{t_{p}} (x q_1q_2 \cdots q_{p-1}), \nonumber \\
\g^2_-(x) & = & \G_{-}^{t_{p+1}} (xq_1q_2\cdots q_p) \cdots \G_{-}^{t_{N+1}} (x q_1q_2 \cdots q_{N}). \nonumber
\eea
With these definitions we can simply write
\be
\overline{W}_p(x) = \g^1_-(x) \g^2_+(x/q) \widehat{Q},    \qquad \qquad   \overline{W}_p'(x) =\g^1_+(x/q) \g^2_-(x) \widehat{Q}.   \label{W-gamma}   
\ee
Moreover $A_{\pm}$ defined in (\ref{Aplus}) and (\ref{Aminus}) can be written (for any $p\in\overline{1,N}$) as
\be
A_{\pm}(x) = \g^1_{\pm}(x) \g^2_{\pm}(x).    \label{Apm-gamma}
\ee

When the argument of any of these operators is $x=1$, we will often use a simplified notation in which this argument is skipped, i.e.
$$
\overline{A}_{\pm} \equiv \overline{A}_{\pm}(1), \qquad A_{\pm} \equiv A_{\pm}(1), \qquad \overline{W}_p \equiv \overline{W}_p(1), \qquad \overline{W}'_p \equiv \overline{W}'_p(1).
$$


\subsection{Quantization of geometry}         \label{ssec-FockState}

In the previous section we considered a toric manifold specified by a triangulation of a strip, and encoded its structure in operators $\overline{A}_{\pm}$, which are specified by a sequence of $\G$ and $\G'$. We can actually do more and represent each such manifold by two states in the Hilbert space of a free fermion $\cH$
$$
|\Omega_{\pm}\rangle  \in\cH.
$$ 
We define these states as
\bea
\langle \Omega_+| & = & \langle 0 | \ldots \overline{A}_+(1) \overline{A}_+(1) \overline{A}_+(1) = \langle 0 | \ldots A_+(q^2) A_+(q) A_+(1),  \label{Omega-plus}  \\
| \Omega_- \rangle & = & \overline{A}_-(1) \overline{A}_-(1) \overline{A}_-(1) \ldots |0\rangle = A_-(1) A_-(q) A_-(q^2) \ldots |0\rangle .  \label{Omega-minus}
\eea

To define these states we used only the classical data of the toric manifold, which is encoded in operators $\overline{A}_+(1)$. Nonetheless they carry information about the full instanton part of the topological string amplitudes, to all orders in string coupling. 
Our first claim is that the overlap of these states
\be
Z = \langle \Omega_+ | \Omega_- \rangle,    \label{Z-Omega}  
\ee
is equal to the BPS partition function $\cZ$ in the chamber corresponding to the non-commutative Donaldson-Thomas invariants
\be
Z = \cZ \equiv |\cZ_{top}|^2 \equiv \cZ_{top}(Q_i) \cZ_{top}(Q_i^{-1}),     \label{Z-cZ}
\ee
with $\cZ_{top}(Q_i)$ given in (\ref{Ztop-strip}), and under the following identification between $q_i$ parameters which enter a definition of $|\Omega_{\pm}\rangle$ and string parameters $Q_i=e^{-T_i}$ and $q_s=e^{-g_s}$:
\be
q_i = (t_i t_{i+1}) Q_i,\qquad \qquad q_s = q \equiv q_0 q_1\cdots q_N.   \label{qQ}
\ee
The proof of the statement (\ref{Z-cZ}) is given in section \ref{ssec-ProofQuantize}.

We note that the states $|\Omega_{\pm}\rangle$ are different than the states $|V\rangle$ which one can associate to the same geometry in the B-model picture of \cite{adkmv}. They have also different properties. In particular, in the framework of \cite{adkmv} the expression of the form $\langle V|V\rangle = \cZ'_{top}$ would represent the topological string partition function of the manifold obtained from gluing two copies of a given manifold to each other. In our case the overlap (\ref{Z-Omega}) gives the square of the topological string partition function of the manifold itself. We also stress that the states $|V\rangle$ are suitable for the large radius limit point in the moduli space, whereas $|\Omega_{\pm}\rangle$ are naturally associated with the non-commutative Donaldson-Thomas chamber. Nonetheless, similarly as for the states $|V\rangle$, it is tempting to think of the states $|\Omega_{\pm}\rangle$ as providing some wave-function interpretation of the underlying classical manifolds, in the spirit of \cite{witten-bckg,top-wave}.

The states $|\Omega_{\pm}\rangle$ relate to the topological string partition function and characterize an extremal chamber in which non-commutative Donaldson-Thomas invariants are defined. There is another extremal case corresponding to the chamber with a single D6-brane with no bound states, so that BPS partition function reads
$$
\widetilde{\cZ} = 1.     
$$
In our formalism in this chamber this partition function can be understood simply as
\be
\widetilde{Z} = \langle 0 | 0 \rangle = 1,    \label{Ztilde-1}
\ee
and clearly $\widetilde{\cZ} = \widetilde{Z}$. This suggests associating the vacuum state $|0\rangle$ to the manifold. We will see shortly that this association makes sense also in multitude of other chambers.


\subsection{Wall-crossing operators}   \label{ssec-WallCross}

In the previous section we realized BPS partition functions in two extreme chambers as fermionic correlators. Now we show that BPS generating functions can be realized as fermionic correlators also in various other chambers.

As discussed in section \ref{ssec-wallcross} only those bound states of D0, D2 or anti-D2-branes with D6-brane can exist for which the central charge (\ref{Zcentral}) is positive
$$
Z(R,B) = \frac{1}{R}(n+\beta \cdot B) > 0.
$$
The fermionic correlators we are after must therefore contain an information about the moduli $R$ and $B$. Our first claim is that the information about $R$ is encoded in the ground state which represents a manifold. This ground state depends only on the sign of $R$ and should be chosen as follows
\be
R>0 \quad \longrightarrow \quad |\Omega_{\pm} \rangle, \qquad \qquad R<0 \quad \longrightarrow \quad |0\rangle.   \label{DTPT-R}
\ee
This choice is of course consistent with (\ref{Z-Omega}) and (\ref{Ztilde-1}).

Our second claim is that crossing the wall of marginal stability corresponds to the insertion of operators $\overline{W}_p$ or $\overline{W}'_p$ defined in (\ref{Wp}) and (\ref{Wpprim}). For this reason we call these operators the \emph{wall-crossing operators}.   Insertion of these operators contains information about the amount of the $B$-field turned on. In this paper we analyze chambers corresponding to an arbitrary flux of the $B$-field through only one, but arbitrary $\mathbb{P}^1$ in the manifold. We fix the number of this $\mathbb{P}^1$ as $p$, see figure \ref{fig-strip}. We claim that insertion of $n$ copies of operators $\overline{W}_p$ or $\overline{W}'_p$ creates respectively $n$ positive or negative quanta of the flux through $p$'th $\mathbb{P}^1$. Therefore, depending on the signs of $R$ and $B$, we have in total four possible situations which we consider separately below. Proofs of all statements in these four situations are given in section \ref{ssec-ProofWalls}.


\subsubsection*{Chambers with $R<0$, $B>0$}


Consider a chamber characterized by positive $R$ and positive $B$-field through $p$'th two-cycle
$$
R<0,\qquad \qquad  B\in ] n-1,n [ \quad \textrm{for}\quad 1\leq n \in \mathbb{Z}.
$$ 
The BPS partition function in this chamber contains only those factors which include $Q_p$ and it reads
$$
\widetilde{\cZ}_{n|p} = \prod_{i=1}^{n-1}  \prod_{s=1}^p \prod_{r=p+1}^{N+1} \Big(1 - \frac{q_s^{i}}{Q_s Q_{s+1}\cdots Q_{r-1}}  \Big) ^{-t_r t_s i}.
$$ 

On the other hand we can compute the expectation value of $n$ wall-crossing operators $\overline{W}_p$. We find that
\be
\widetilde{Z}_{n|p}  =  \langle 0| ( \overline{W}_p ) ^n |0 \rangle   = 
\prod_{i=1}^{n-1}  \prod_{s=1}^p \prod_{r=p+1}^{N+1} \Big(1 - (t_rt_s)\frac{q^{n-i}}{q_s q_{s+1}\cdots q_{r-1}}  \Big) ^{-t_r t_s i}.   \label{ZRnegBpos-Results}  \\
\ee
The proof of this equality is given is section \ref{ssec-ProofWalls}.

Therefore, under the following change of variables
\be
Q_p = (t_p t_{p+1})q_p q_s^{n},\qquad \qquad Q_i = (t_i t_{i+1})q_i \quad \textrm{for}\quad i\neq p,\qquad \qquad q_s = \frac{1}{q}.  \label{ZRnegBpos-qQ-Results}
\ee
the correlator (\ref{ZRnegBpos-Results}) reproduces the BPS partition function 
\be
\widetilde{\cZ}_{n|p} = \widetilde{Z}_{n|p}.       \label{ZRnegBpos-equal}
\ee

An insertion of $\overline{W}_p$ has an interpretation of turning on a positive quantum of $B$-field, and the redefinition of $Q_p$ can be interpreted as effectively enlarging K{\"ahler} parameter $T_p$ by one unit of $g_s$. In fact the minimal number of insertions is $n=1$. Because in $\overline{W}_p$ all $\G_+$ operators are to the right of $\G_-$, an insertion of one operator does not have any effect, so 
$$
\widetilde{Z}_{1|p} = \langle 0| \overline{W}_p |0 \rangle = \langle 0|0 \rangle = 1,
$$
still represents a chamber with a single D6-brane and no other branes bound to it.


\subsubsection*{Chambers with $R>0$, $B>0$}


In the second case we consider the positive value of $R$ and the positive flux through $p$'th $\mathbb{P}^1$
$$
R>0, \qquad \qquad B\in ]n,n+1 [ \quad \textrm{for} \quad   0\leq n \in \mathbb{Z}.
$$ 
Denote the BPS partition function in this chamber by $\cZ_{n|p}$.

We find that the expectation value of $n$ wall-crossing operators $\overline{W}_p$ in the background of $|\Omega\rangle$ has the form 
\be
Z_{n|p} = \langle \Omega_+| ( \overline{W}_p ) ^n | \Omega_-\rangle  = M(1,q)^{N+1} \, Z^{(0)}_{n|p} \, Z^{(1)}_{n|p} \, Z^{(2)}_{n|p},  \label{ZRposBpos-Results}
\ee
where $Z^{(0)}_{n|p}$ does not contain any factors $(q_s\cdots q_{r-1})^{\pm 1}$ which would include $q_p$, while $Z^{(1)}_{n|p}$ contains all factors $q_s\cdots q_{r-1}$ which do include $q_p$, and $Z^{(2)}_{n|p}$ contains all factors $(q_s\cdots q_{r-1})^{-1}$ which also include  $q_p$:
\bea
Z^{(0)}_{n|p} & = & \prod_{l=1}^{\infty} \prod_{p \notin \overline{s,r+1} \subset \overline{1,N+1}} \Big(1 - (t_rt_s) \frac{q^{l}}{ q_s q_{s+1}\cdots q_{r-1}}  \Big) ^{-t_r t_s l}  \Big(1 - (t_rt_s) q^{l} q_s q_{s+1}\cdots q_{r-1} \Big) ^{-t_r t_s l} ,    \nonumber \\
Z^{(1)}_{n|p} & = & \prod_{l=1}^{\infty} \prod_{p \in \overline{s,r+1} \subset \overline{1,N+1}}   \Big(1 - (t_rt_s) q^{l+n} q_s q_{s+1}\cdots q_{r-1} \Big) ^{-t_r t_s l} ,    \nonumber \\
Z^{(2)}_{n|p} & = & \prod_{l=n+1}^{\infty} \prod_{p \in \overline{s,r+1} \subset \overline{1,N+1}} \Big(1 - (t_rt_s) \frac{q^{l-n}}{ q_s q_{s+1}\cdots q_{r-1}}  \Big) ^{-t_r t_s l} .     \nonumber 
\eea

Clearly the change of variables
\be
Q_p = (t_p t_{p+1})q_p q_s^{n},\qquad \qquad Q_i = (t_i t_{i+1})q_i \quad \textrm{for}\quad i\neq p,\qquad \qquad q_s = q  \label{ZRposBpos-qQ-Results}
\ee
reproduces the BPS partition function
\be
\cZ_{n|p} = Z_{n|p}.             \label{ZRposBpos-equal}
\ee
When no wall-crossing operator is inserted the change of variables reduces to (\ref{qQ}) and we get the non-commutative Donaldson-Thomas partition function (\ref{Z-cZ}), $Z_{0|p}=\cZ$.

The proof of (\ref{ZRposBpos-Results}) and in consequence (\ref{ZRposBpos-equal}) is given in section \ref{ssec-ProofWalls}.


\subsubsection*{Chambers with $R<0$, $B<0$}


Now we consider negative $R$ and negative $B$-field 
$$
R<0, \qquad \qquad   B \in ]-n-1,-n [ \quad \textrm{for} \quad 0\leq n \in \mathbb{Z}.
$$
For such a chamber the BPS partition function reads
$$
\widetilde{\cZ}'_{n|p} = \prod_{i=1}^{n}  \prod_{s=1}^p \prod_{r=p+1}^{N+1} \Big(1 - q_s^{i} Q_s Q_{s+1}\cdots Q_{r-1}  \Big) ^{-t_r t_s i}.
$$

Now we find the the expectation value of $n$ wall-crossing operators $\overline{W}'_p$ is equal to
\be
\widetilde{Z}'_{n|p} = \langle 0| ( \overline{W}'_p ) ^n | 0 \rangle  
= \prod_{i=1}^{n}  \prod_{s=1}^p \prod_{r=p+1}^{N+1} \Big(1 - (t_rt_s) q^{n-i} q_s q_{s+1}\cdots q_{r-1}  \Big) ^{-t_r t_s i}.  \label{ZRnegBneg-Results}
\ee
This equality is proved in section \ref{ssec-ProofWalls}. Therefore, under a change of variables
\be
Q_p = (t_p t_{p+1})q_p q_s^{-n},\qquad \qquad Q_i = (t_i t_{i+1})q_i \quad \textrm{for}\quad i\neq p,\qquad \qquad q_s = \frac{1}{q},  \label{ZRnegBneg-qQ-Results}
\ee
this reproduces the BPS partition function 
\be
\widetilde{\cZ}'_{n|p} = \widetilde{Z}'_{n|p}.   \label{ZRnegBneg-equal}
\ee

Now an insertion of $\overline{W}_p$ has an interpretation of turning on a negative quantum of $B$-field, and the redefinition of $Q_p$ can be interpreted as effectively reducing $t_p$ by one unit of $g_s$. As we already discussed,
$$
\widetilde{Z}'_{0|p} = \langle 0 | 0 \rangle =  1
$$
represents a chamber with a single D6-brane and no other branes bound to it. Contrary to the case with $B>0$, an insertion of a single $\overline{W}'_p$ has a non-trivial effect.


\subsubsection*{Chambers with $R>0$, $B<0$}


In the last case we consider positive $R$ and negative $B$
$$
R>0, \qquad \qquad  0>B\in ]-n,-n+1 [ \quad \textrm{for} \quad   1\leq n \in \mathbb{Z}.
$$
We denote the BPS partition function in this chamber by $\cZ'_{n|p}$.

We find that the expectation value of $n$ operators $\overline{W}'_p$ in the background of $|\Omega_{\pm}\rangle$ has the form
\be
Z'_{n|p} = \langle \Omega_+| ( \overline{W}'_p ) ^n | \Omega_-\rangle  =   M(1,q)^{N+1} \, Z'^{(0)}_{n|p} \, Z'^{(1)}_{n|p} \, Z'^{(2)}_{n|p}  \label{ZRposBneg-Results}
\ee
where $Z'^{(0)}_{n|p}$ does not contain any factors $(q_s\cdots q_{r-1})^{\pm 1}$ which would include $q_p$,  $Z'^{(1)}_{n|p}$ contains all factors $q_s\cdots q_{r-1}$ which do include $q_p$, and $Z'^{(2)}_{n|p}$ contains all factors $(q_s\cdots q_{r-1})^{-1}$ which also include  $q_p$:
\bea
Z'^{(0)}_{n|p} & = & \prod_{l=1}^{\infty} \prod_{p \notin \overline{s,r+1} \subset \overline{1,N+1}} \Big(1 - (t_rt_s) \frac{q^{l}}{ q_s q_{s+1}\cdots q_{r-1}}  \Big) ^{-t_r t_s l}  \Big(1 - (t_rt_s) q^{l} q_s q_{s+1}\cdots q_{r-1} \Big) ^{-t_r t_s l} ,    \nonumber \\
Z'^{(1)}_{n|p} & = & \prod_{l=n}^{\infty} \prod_{p \in \overline{s,r+1} \subset \overline{1,N+1}}   \Big(1 - (t_rt_s) q^{l-n} q_s q_{s+1}\cdots q_{r-1} \Big) ^{-t_r t_s l} ,    \nonumber \\
Z'^{(2)}_{n|p} & = & \prod_{l=1}^{\infty} \prod_{p \in \overline{s,r+1} \subset \overline{1,N+1}} \Big(1 - (t_rt_s) \frac{q^{l+n}}{ q_s q_{s+1}\cdots q_{r-1}}  \Big) ^{-t_r t_s l} .     \nonumber 
\eea

Clearly the change of variables
\be
Q_p = (t_p t_{p+1})q_pq_s^{-n-1},\qquad \qquad Q_i = (t_i t_{i+1})q_i \quad \textrm{for}\quad i\neq p,\qquad \qquad q_s = q.  \label{ZRposBneg-qQ-Results}
\ee
reproduces the BPS partition function
\be
\cZ'_{n|p} = Z'_{n|p}.    \label{ZRposBneg-equal}
\ee

We note that both $Z'_{1|p}$ with the above change of variables, as well as $Z_{0|p}$ given in (\ref{ZRposBpos-Results}) with a different change of variables in (\ref{ZRposBpos-qQ-Results}), lead to the same BPS generating function $\cZ$ which corresponds to the non-commutative Donaldson-Thomas invariants.

The proof of (\ref{ZRposBneg-Results}) and in consequence (\ref{ZRposBneg-equal}) is given in section \ref{ssec-ProofWalls}.


\subsection{Crystal melting interpretation}   \label{ssec-CrystalMelting}

So far we have found a representation of D6-D2-D0 generating functions as correlators in the free fermion theory. 
In this section we discuss crystal melting interpretation of these correlators, and explain how to associate crystal models to local, toric manifolds without compact four-cycles, based on the results described above. Our point of view generalizes crystal models found previously, such as plane partitions for $\mathbb{C}^3$ crystal summarized briefly in section \ref{ssec-crystal} or pyramid partitions for the conifold crystal from section \ref{ssec-ReviewConifold}, and provides an interesting unifying perspective. Furthermore we discuss evolution of crystals upon changing the moduli of the theory.


Moreover, for the class of manifolds which we consider in this paper, we claim that our crystals are equivalent to colored crystals introduced in \cite{Ooguri-crystal} in terms of quiver diagrams and relation to dimers. 
The construction of crystals explained in this section and their evolution upon wall-crossing is illustrated in several examples in section \ref{sec-examples}.

\subsubsection*{Construction of crystals}

Our crystal interpretation is inherently related to the form of fermionic correlators which we found in sections \ref{ssec-FockState} and \ref{ssec-WallCross}. All these correlators are constructed from operators $\overline{A}_{\pm}$, $\overline{W}_p$ and $\overline{W}'_p$, which involve only vertex operators $\G_{\pm}$ and $\G_{\pm}$ with argument 1 and color operators $\widehat{Q}_i$. According to the relations (\ref{Gmu}) and (\ref{Gprimmu}), insertion of these vertex operators can be interpreted as insertion of two-dimensional partitions satisfying interlacing, or transposed interlacing conditions. This corresponds to constructing the three-dimensional crystal from two-dimensional slices. A relative position of the neighboring slices is determined by which vertex operator they are created. Additional insertions of color operators have an interpretation of coloring the crystal. Because in all operators which we consider in this paper the colors $\widehat{Q}_i$ appear in the same order, these colors are always repeated periodically in the full correlators, and in consequence our crystals are made of interlacing periodically colored slices.

\begin{figure}[htb]
\begin{center}
\includegraphics[width=0.5\textwidth]{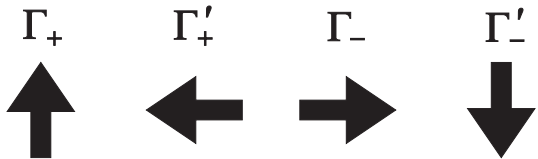} 
\begin{quote}
\caption{\emph{Assignment of arrows.}} \label{fig-arrows}
\end{quote}
\end{center}
\end{figure}

All the information about the crystal, including its shape, coloring and interlacing pattern, can be encoded in a simple graphical form. To do this we associate various arrows to the vertex operators. This assignment is shown in figure \ref{fig-arrows}. The arrows are always drawn from left to right, or up to down (a direction of drawing is independent of the orientation of the arrow). Then we translate a sequence of vertex operators which appear in a given correlator into a sequence of corresponding arrows, and draw them such that the end of one arrow becomes the end of the next one. We also keep track of the coloring by drawing at the endpoint of each arrow a (dashed) line, rotated by 45$^\textrm{o}$, colored according to $\widehat{Q}_i$ which we come across. These lines represent two-dimensional slices in appropriate colors.

The zig-zag path which arises from the above prescription represents the shape of the crystal, as seen from the top. In particular, the corners of two-dimensional partitions arising from slicing of the crystal are located at the end-points of the arrows. The orientation of arrows represents the interlacing condition (i.e. arrows point from a larger to smaller partition), and therefore it indicates a direction in which the crystal grows. The interlacing pattern between two consecutive slices corresponds to the types of two consecutive arrows. 
Finally, the points from which two arrows point outwards represent those stones in the crystal, which can be removed from the initial, full crystal configuration. In fermionic correlators these points correspond to $\G^{t_i}_+$ followed by $\G^{t_j}_-$ operators. All these statements are easy to check in the examples presented in section \ref{sec-examples}.

\subsubsection*{Evolution of crystals}   \label{sssec-evolution}

It is also interesting to analyze the evolution of crystals upon crossing walls of marginal stability. 
We stress that by a fixed crystal we understand a set of all admissible configurations of its constituents (such as e.g. plane partitions) which fit into a \emph{fixed} container (such as the positive octant of $\mathbb{C}^3$). The term \emph{evolution} refers to the evolution of a shape of this underlying container, upon which the set of admissible crystal configurations of course changes too. Such an evolution arises from changing values of moduli, so that the values of the central charge (\ref{Zcentral})
$$
Z(R,B) = \frac{1}{R}(l + B\cdot \beta)  
$$
also change. 

We consider first increasing or decreasing $B$-field through a fixed $p$'th $\mathbb{P}^1$. 
We focus first on the noncommutative Donaldson-Thomas chamber with the BPS partition functions given by (\ref{Z-Omega})
$$
Z = \langle \Omega_+ | \Omega_- \rangle = \langle 0| \ldots \overline{A}_+ \overline{A}_+ \overline{A}_+ | \overline{A}_- \overline{A}_- \overline{A}_- \ldots|0 \rangle.
$$
As all $\overline{A}_{\pm}$ operators are built from $\G^{t_i}_{\pm}$ vertex operators, the crystal for this chamber is built in increasing direction following the string of $\overline{A}_+$'s, and then in decreasing direction following the string of $\overline{A}_-$'s. There is only one point from which two arrows point outwards, or equivalently $\G_+$ is followed by $\G_-$: this is just the point when $\overline{A}_+$ turns into $\overline{A}_-$ in the above correlator. Therefore for each manifold in this extreme chamber the crystal has only one corner which can be removed from the initial, fully filled crystal configuration.

Moving to other chambers corresponds to changing the $B$-field and inserting wall-operators, and therefore the structure of the crystal gets deformed. As explained in section \ref{ssec-triangulate}, the wall-crossing operators (\ref{Wp}) and (\ref{Wpprim}) consist of a string of $\G^{t_i}_{+}$'s followed by a string of $\G^{t_j}_{-}$, or vice versa. Therefore insertion of each such wall-crossing operator introduces one additional corner of the crystal. One exemption from this rule arises when $\overline{W}'_p$ is inserted into the extreme correlator (\ref{Z-Omega}): then the single vertex from the extreme configuration is replaced by a new single vertex encoded in $\overline{W}'_p$.

One can also start from the other extreme chamber corresponding to a single D6-brane (\ref{Ztilde-1}). Turning on the $B$-field also modifies the crystal. An insertion of the first $\overline{W}_p$ operator has no effect, as it has all $\G^{t_i}_+$ to the right of $\G^{t_j}_-$. Therefore a chamber characterized by an insertion of $(\overline{W}_p)^{n+1}$ corresponds to a crystal with $n$ corners. Similarly, a chamber characterized by an insertion of $(\overline{W}'_p)^{n}$ corresponds to a crystal with $n$ corners.

We can also consider changing the sign of the $B$-field. This corresponds to changing the counting of M2-branes into counting of the anti-M2-branes. At least for the conifold, this can also be interpreted as a flop transition. To change a sign one has to remove all $\overline{W}_p$ operators, cross the extreme chamber, and then start adding $\overline{W}'_p$ operators (or vice versa). From the assignment of arrows in figure \ref{fig-arrows} we observe that changing a type of wall-crossing operator corresponds to a perpendicular change of the direction in which crystal is expanding. Such a crystal interpretation of the flop transition in the conifold case will be discussed in section \ref{ssec-flop}.

Finally one can consider changing a sign of $R$. This corresponds to changing the ground state representing the manifold according to (\ref{DTPT-R}). For $R>0$ the ground state is represented by $|\Omega_{\pm}\rangle$ which includes of infinite number of $\overline{A}_{\pm}$ operators, and therefore the crystal extends infinitely in both associated directions. When $R$ changes its sign to negative values the ground state gets replaced by $|0\rangle$, and crystal becomes finite in both associated directions, with the size specified by the number of wall-crossing operators inserted. This dramatic change of size of the crystal provides an interpretation of the so-called DT/PT transition \cite{MNOP-I,DTPT}. Moreover, in some cases, such as the conifold discussed in sections \ref{ssec-ReviewConifold} and \ref{ssec-ExampleConifold}, the whole crystal is finite for $R<0$. However the crystal can also extend infinitely in the third dimension, as is the case for orbifold of $\mathbb{C}^3$ discussed in section \ref{ssec-OrbiC3}.

The above discussion focused on $\overline{W}_p$ and $\overline{W}'_p$ operators. It should be possible to generalize them to account for all possible chambers, i.e. construct operators which would represent arbitrary $B$-fields through arbitrary set of $\mathbb{P}^1$'s, not just one fixed $\mathbb{P}^1$. Examples of such more general operators will be discussed in section \ref{ssec-triple}. In those more general cases the evolution of crystals is of course more complicated.


\section{Examples}    \label{sec-examples}

In this section we illustrate the results presented in section \ref{sec-results} in several instructive examples, which include orbifolds of $\mathbb{C}^3$, resolved conifold and triple-$\mathbb{P}^1$ geometry. We discuss a crystal interpretation of the flop transition for the conifold. Finally we find a crystal description of the part of the chamber space of the closed topological vertex, taking advantage of its relation to the triple-$\mathbb{P}^1$ geometry and physical viewpoint from section \ref{ssec-wallcross}.

\subsection{Revisiting $\mathbb{C}^3$}    \label{ssec-C3}

\begin{figure}[htb]
\begin{center}
\includegraphics[width=0.8\textwidth]{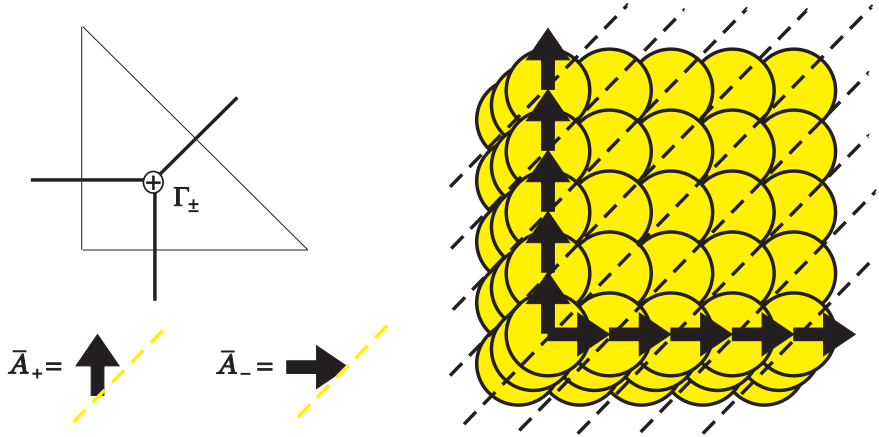} 
\begin{quote}
\caption{\emph{Toric diagram for $\mathbb{C}^3$ (upper left) consists of one triangle and has one $\oplus$ vertex. Therefore $\overline{A}_{\pm}$ operators involve just single $\G_{\pm}$, which are represented by arrows (lower left) according to figure \ref{fig-arrows}. The correlator (\ref{C3-crystal}) is translated into a sequence of arrows, with rotated dashed lines representing insertions of interlacing two-dimensional partitions. The resulting figure (right) represents MacMahon crystal for plane partitions from figure \ref{fig-C3}, seen from the bottom (first three layers are shown; it is assumed that each stone in layer $m+1$ is blocked by only one stone located immediately above it in layer $m$).}} \label{fig-C3new}
\end{quote}
\end{center}
\end{figure}

To start with we reconsider the simplest case of $\mathbb{C}^3$ crystal and explain how it fits into the prescription from section \ref{sec-results}. For $\mathbb{C}^3$ a triangulation of a strip consists just of one triangle of area $1/2$, see figure \ref{fig-C3new} (left). Therefore there is just one vertex and only one color $\widehat{Q}_0\equiv \widehat{Q}$. The operators (\ref{Apm}) take the form
$$
\overline{A}_{\pm} = \G_{\pm}(1) \widehat{Q},
$$
and there are no wall-crossing operators. Therefore the BPS partition function (\ref{Z-Omega}) takes exactly the form (\ref{C3-crystal}) and we find that the BPS generating function is given by the MacMahon function. 

To reconstruct the crystal we associate arrows to $\widehat{A}_{\pm}$ operators according to figure \ref{fig-arrows}, and draw them in the order which follows the order of vertex operators in (\ref{Z-Omega}). The crystal which we obtain is shown in figure \ref{fig-C3new} (right). We indeed reproduce plane partitions from section \ref{ssec-crystal}, which in our picture are seen from the other side than in the more often encountered figure \ref{fig-C3}. 

Even though there are no wall-crossing operators in this case, there is the other extreme chamber with a single D6-brane, for which the generating function is given by (\ref{Ztilde-1}).

\subsection{Orbifolds $\mathbb{C}^3 / \mathbb{Z}_{N+1}$}    \label{ssec-OrbiC3}

\begin{figure}[htb]
\begin{center}
\includegraphics[width=\textwidth]{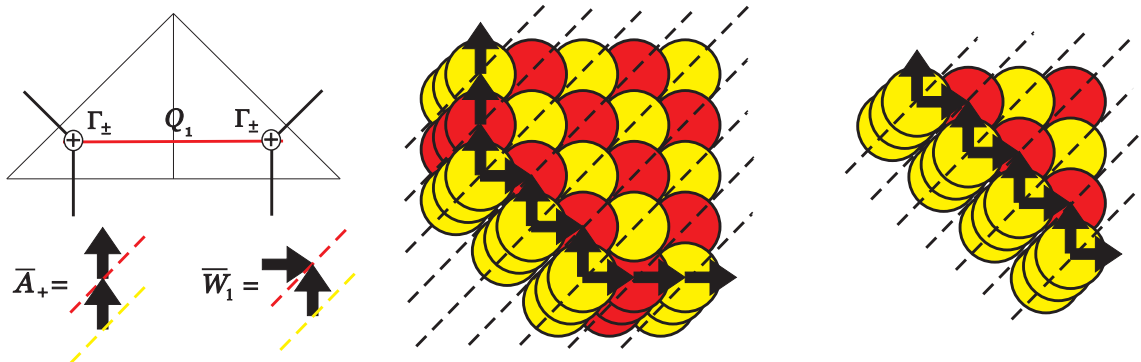} 
\begin{quote}
\caption{\emph{Toric diagram for the resolution of $\mathbb{C}^3/ \mathbb{Z}_{N+1}$ geometry has $N+1$ vertices of the same type $\oplus$; this figure shows the case $N=1$. Left: toric diagram and translation of $\overline{A}_+$ and $\overline{W}_1$ into arrows. In the non-commutative Donaldson-Thomas chamber this leads to the same plane partition crystal as in figure \ref{fig-C3new}, however colored now in yellow and red. Middle: for the chamber with positive $R$ and $2<B<3$ the crystal develops two additional corners and its partition function is given by $Z_{2|1} = \langle \Omega_+|(\overline{W}_1)^2|\Omega_-\rangle$. Right: for negative $R$ and positive $n-1<B<n$ the crystal is finite along two axes (albeit still infinite along the third axis perpendicular to the picture) and develops $n-1$ yellow corners; its generating function for the case of $n=5$ shown in the picture reads $\widetilde{Z}_{5|1} = \langle 0 |(\overline{W}_1)^5|0\rangle$ (two external arrows, corresponding to $\G_-$ acting on $\langle 0|$ and $\G_+$ acting on $|0\rangle$, are suppressed.)}} \label{fig-OrbiC3}
\end{quote}
\end{center}
\end{figure}

We apply now the prescription from section \ref{sec-results} to the case of (the resolution of) $\mathbb{C}^3 / \mathbb{Z}_{N+1}$ orbifold. The toric diagram looks like a big triangle of area $(N+1)/2$. There are $N$ independent $\mathbb{P}^1$'s, as well as $N+1$ vertices of the same $t_i=+1$ type, see figure \ref{fig-OrbiC3} (left). Therefore operators in (\ref{Apm}) take the form
$$
\overline{A}_{\pm} = \G_{\pm}(1) \widehat{Q}_1 \G_{\pm}(1) \widehat{Q}_2 \ldots \G_{\pm}(1) \widehat{Q}_{N} \G_{\pm}(1) \widehat{Q}_0.
$$
Thus, in the non-commutative Donaldson-Thomas chamber, the corresponding crystal consists of plane partitions just as in figure \ref{fig-C3new}, however now with periodically colored slices in $N+1$ colors. The non-commutative Donaldson-Thomas partition function is given by (\ref{Z-Omega}). This reproduces the results for $\mathbb{C}^3 / \mathbb{Z}_{N+1}$ orbifolds from \cite{YoungBryan}.

We can also analyze wall-crossing related to turning on arbitrary $B$-field through a fixed $\mathbb{P}^1$. Translating the wall-crossing operators (\ref{Wp}) and (\ref{Wpprim}) into arrows following figure \ref{fig-arrows}, we obtain crystals in modified containers, as shown in the example in figure \ref{fig-OrbiC3} (middle). Enlarging the $B$-field by one unit, which corresponds to an insertion of one wall-operator $\overline{W}_p$, adds one more yellow corner to the crystal. Applying wall-crossing operators $\overline{W}'_p$ would result in a crystal which develops red corners.

We also immediately find crystals corresponding to $R<0$. As usual the crystal is empty in the extreme chamber with a single D6-brane (\ref{Ztilde-1}). Adding wall-crossing operators results in a crystal which develops corners, as shown in figure \ref{fig-OrbiC3} (right). This crystal is finite along two axes. Nonetheless, because $\overline{W}_p$ and $\overline{W}'_p$ consist only of $\G_{\pm}$ operators (and no $\G'_{\pm}$ are involved), it can grow infinitely along the third axis. 


\subsection{Resolved conifold}    \label{ssec-ExampleConifold}

We reviewed wall-crossing for the conifold in section \ref{ssec-ReviewConifold}. We show now that analyzing it from a perspective of section \ref{sec-results} provides a new proof of some statements posed in literature and extends them to a wide class of manifolds.


\begin{figure}[htb]
\begin{center}
\includegraphics[width=0.9\textwidth]{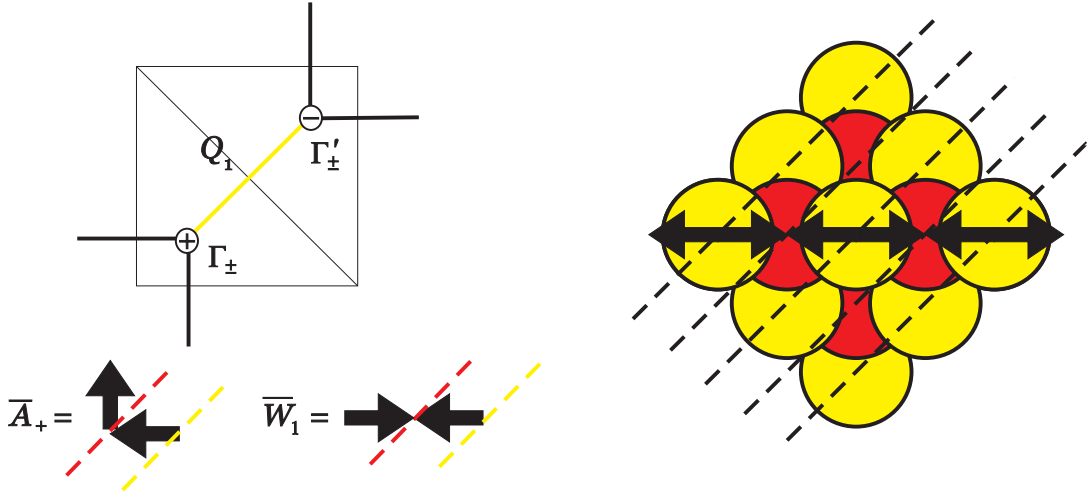} 
\begin{quote}
\caption{\emph{Left: toric diagram for the conifold and translation of $\overline{A}_+$ and $\overline{W}_1$ operators into arrows. Right: for chambers with negative $R$ and positive $n-1<B<n$ the crystals are given by finite pyramid partitions with $n-1$ additional corners developed, represented by $n-1$ stones in the top row. The corresponding partition function is given by $\widetilde{Z}_{n|1} = \langle 0|(\overline{W}_1)^n|0\rangle$ which reproduces the result (\ref{ZZn}) reviewed in section \ref{ssec-ReviewConifold}. This figure shows the case $n=4$ (again two external arrows representing $\langle 0|\G_-$ and $\G'_+|0\rangle$ are suppressed).}} \label{fig-pyramid-coni-inf3}
\end{quote}
\end{center}
\end{figure}

The toric diagram for the conifold is shown in figure \ref{fig-pyramid-coni-inf3} (left). There is of course just $N=1$ $\mathbb{P}^1$, 
and two colors $\widehat{Q}_1$ and $\widehat{Q}_0$, so that
$$
\widehat{Q} = \widehat{Q}_1 \widehat{Q}_0, \qquad \qquad q=q_1 q_0.
$$
The operators (\ref{Apm}) take form
$$
\overline{A}_{\pm}(x) = \G_{\pm}(x) \widehat{Q}_1 \G'_{\pm}(x) \widehat{Q}_0,
$$
while (\ref{Aplus}) and (\ref{Aminus}) read
$$
A_+(x) = 
\G_+(xq ) \G'_+(x q / q_1), \qquad \qquad A_-(x) = 
\G_-(x) \G'_-(xq_1),
$$
and satisfy
$$
A_+(x) A_-(y) = \frac{(1+xy q / q_1)(1+xy q q_1)}{(1-xy q)^2}A_-(y) A_+(x).
$$

The quantum geometry is encoded in quantum states (\ref{Omega-plus}) and (\ref{Omega-minus})
\bea
|\Omega_-\rangle & = & 
A_-(1) A_-(q) A_-(q^2)\ldots |0\rangle,  \\
\langle \Omega_+ | & = & 
\langle 0| \ldots A_+(q^2) A_+(q) A_+(1) 
\eea
and the wall-insertion operators (\ref{Wp}) (\ref{Wpprim}) are
\bea
\overline{W}_1(x) & = & \G_-(x) \widehat{Q}_1 \G'_+(x) \widehat{Q}_0, \\
\overline{W}'_1(x) & = & \G_+(x) \widehat{Q}_1 \G'_-(x) \widehat{Q}_0.     \label{Wp-conifold}
\eea

\begin{figure}[htb]
\begin{center}
\includegraphics[width=0.8\textwidth]{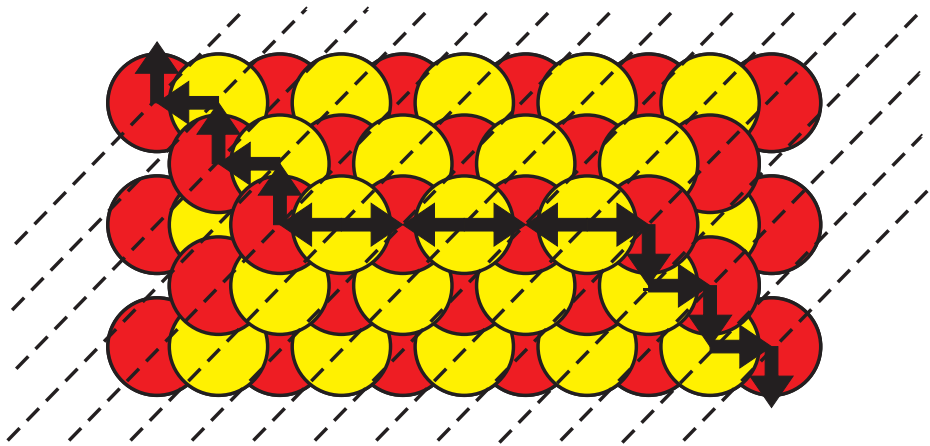} 
\begin{quote}
\caption{\emph{Conifold crystal in the chamber with positive $R$ and $2<B<3$ takes form of pyramid partitions with $3$ stones in the top row. Its generating function is given by $Z_{2|1} = \langle \Omega_+|(\overline{W}_1)^2|\Omega_-\rangle$.}} \label{fig-conifold-crystal}
\end{quote}
\end{center}
\end{figure}



From our results in section (\ref{sec-results}) we immediately find that the fermionic correlators
\bea
Z_{n|1} & = & \langle \Omega_+ |(\overline{W}_1)^n| \Omega_- \rangle,    \\
\widetilde{Z}_{n|1} & = & \langle 0|(\overline{W}_1)^{n} |0 \rangle.   \label{coni-finite-crystal}
\eea
result in the pyramid partition functions (\ref{Zn}) and (\ref{ZZn}) reviewed in section \ref{ssec-ReviewConifold}. 
With no wall-crossing operators inserted the above partition functions give respectively Szendroi's result $Z_{0|1}=\langle\Omega_+|\Omega_-\rangle$ and pure D6-brane $\widetilde{Z}=\langle 0|0\rangle =1$. 

The above correlators have crystal interpretation shown respectively in figures \ref{fig-conifold-crystal} and \ref{fig-pyramid-coni-inf3} (right). These are of course the same pyramid partitions that appeared in earlier literature, see figures \ref{fig-pyramids-coni-inf} and \ref{fig-pyramids-coni-fin}.
Moreover, insertion of the companion $\overline{W}'_1$ operators leads to pyramids that extend in perpendicular directions, as we will discuss in the next section.

In particular, in our formalism we can automatically provide a new proof of a conjecture posed in \cite{ChuangJaff},\footnote{This has been proved by other means by mathematicians in \cite{NagaoNakajima}, where also the geometrical meaning of the quivers constructed in \cite{ChuangJaff} was clarified. We note that our approach shows how to generalize the notion of finite pyramids to the wide class of manifolds considered in this paper.} which states that BPS partition functions in chambers separated by finite numbers of walls from the pure D6-brane region are given by generating functions of finite pyramids. This fact is just a consequence of the existence of the fermionic representation (\ref{coni-finite-crystal}): on one hand we already discussed that this fermionic representation reproduces appropriate generating functions (\ref{ZZn}). On the other hand this representation provides a crystal construction encoded in the arrow structure, as we explained in section \ref{ssec-CrystalMelting}. Contrary to $\mathbb{C}^3 / \mathbb{Z}^{N+1}$ cases, now the crystal pyramids are finite in all three directions, as shown in figure \ref{fig-pyramid-coni-inf3} (right). This is so, because now the wall-operators operators (\ref{Wp-conifold}) involve both $\G$ and $\G'$ operators, which insert interlacing  two-dimensional partitions that extend in opposite directions and effectively block each other beyond the length given by the number of inserted $\overline{W}_1$'s.


\subsection{Flop transition}     \label{ssec-flop}

In section \ref{ssec-CrystalMelting} we discussed evolution of crystals upon changing the moduli $R$ and $B$. It is in particular interesting to focus on the case when the $B$-field which changes the sign. According to the interpretation reviewed in section \ref{ssec-wallcross} this corresponds to the counting of anti-M2-branes instead of M2-branes. However, in case of the conifold such a process can be identified with the flop transition. 

\begin{figure}[htb]
\begin{center}
\includegraphics[width=0.70\textwidth]{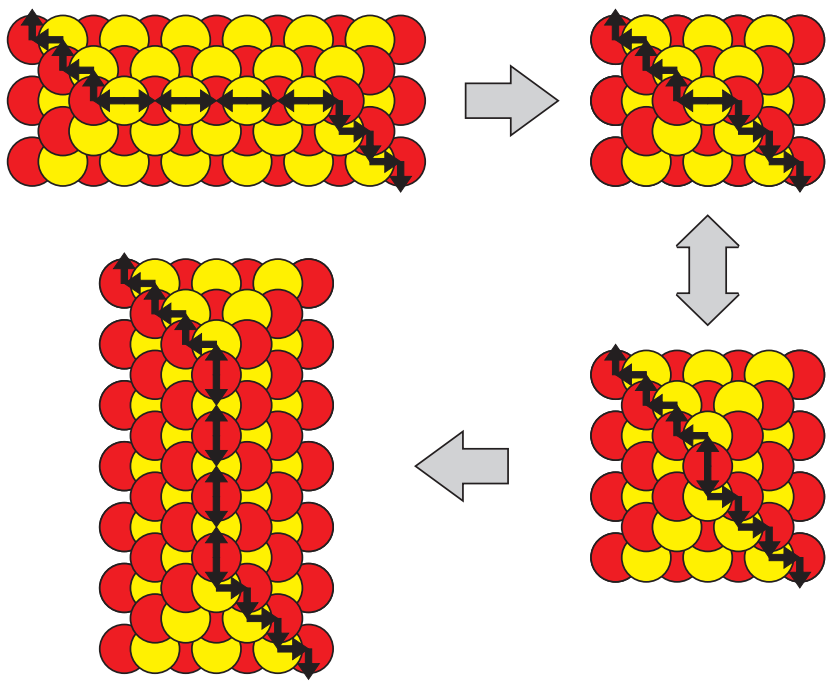} 
\begin{quote}
\caption{\emph{Flop transition of the conifold. Right: there are two equivalent representations of the non-commutative Donaldson-Thomas chamber, with yellow or red stone on top and changes of variables given respectively by (\ref{ZRposBpos-qQ-Results}) and (\ref{ZRposBneg-qQ-Results}). Upper and lower left: extension of the pyramid crystal in opposite directions for positive and negative $B$-field, represented respectively by insertions of $\overline{W}_{1}$ and $\overline{W}'_{1}$ operators. }} \label{fig-flop}
\end{quote}
\end{center}
\end{figure}

The crystal interpretation of the flop transition is shown in figure \ref{fig-flop}. The crystal representing a chamber separated by $n$ walls of marginal stability from the non-commutative Donaldson-Thomas chamber, with K{\"a}hler parameter 
\be
Q_1=-q_1 q_s^{n}      \label{qQ-flop1}
\ee
and the partition function $Z_{n|1} = \langle \Omega_+|(\overline{W}_1)^n|\Omega_-\rangle$, is shown in upper left. The non-commutative Donaldson-Thomas chamber (right) is equivalently represented by $Z_{0|1} = \langle \Omega_+|\Omega_-\rangle\equiv \cZ$ 
(with yellow stone on top and a change of variables $Q_1=-q_1$ in (\ref{ZRposBpos-qQ-Results})), and $Z'_{1|1} = \langle \Omega_+|\overline{W}'_1|\Omega_-\rangle \equiv \cZ$ (with red stone on top and a change of variables $Q_1=-q_1 q^{-1}$ in (\ref{ZRposBneg-qQ-Results})). Further insertions of $\overline{W}'_{1}$ operators (lower left) extend the crystal in perpendicular direction, with the K{\"a}hler parameter identified as
\be
Q_1=-\frac{q_1}{q} q_s^{n}    \label{qQ-flop2}
\ee
and the partition function given by $Z'_{n|1} = \langle \Omega_+|(\overline{W}'_1)^n|\Omega_-\rangle$. Note that changes of variables in (\ref{qQ-flop1}) and (\ref{qQ-flop2}) can be interpreted respectively as increasing and decreasing the K{\"a}hler parameter of the singular conifold $T_1 = -\log Q_1$ by $n$ units of $(-g_s)=\log q_s$. 

Similar picture holds for finite pyramids in chambers with $R<0$, albeit in this case the singularity is represented by the empty pyramid.

\subsection{Triple-$\mathbb{P}^1$ geometry and more general wall-operators}   \label{ssec-triple}

\begin{figure}[htb]
\begin{center}
\includegraphics[width=0.9\textwidth]{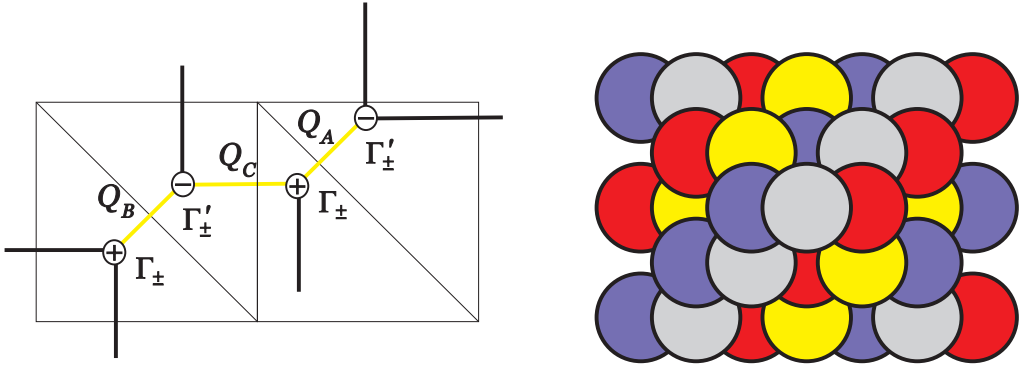} 
\begin{quote}
\caption{\emph{Toric diagram for the triple-$\mathbb{P}^1$ geometry (left) and the corresponding four-colored pyramid crystal in the non-commutative Donaldson-Thomas chamber (we use here subscripts $b,c,a$ to denote various $\mathbb{P}^1$'s instead of $1,2,3$ used earlier).}} \label{fig-triple-P1}
\end{quote}
\end{center}
\end{figure}

We discuss now the triple-$\mathbb{P}^1$ geometry whose toric diagram is shown in figure \ref{fig-triple-P1} (left). While this appears to be an obvious generalization of our previous examples, there are two important reasons to consider this case. Firstly, for this geometry (or any other geometry with longer but analogous toric diagram) a simple geometric intuition allows to introduce more general wall-crossing operators (related to turning on non-trivial $B$-field through all three $\mathbb{P}^1$'s). Secondly, this case is related to the closed topological vertex geometry, which we will discuss in the next section. 

We label\footnote{In this section we change a notation slightly: we label three $\mathbb{P}^1$'s by $B,C,A$ instead of 1,2,3 used in section \ref{sec-results}, and denote corresponding fermionic parameters respectively by $b,c,a$.} the three independent $\mathbb{P}^1$'s in this geometry by $B,C,A$, denote by $T_B, T_C, T_A$ their K{\"a}hler parameters, and introduce
$$
Q_A = e^{-T_A}, \qquad  Q_B = e^{-T_B}, \qquad  Q_C = e^{-T_C}.
$$
The topological string partition function for this geometry reads
\be
\cZ^{triple}_{top}(Q_A,Q_B,Q_C) = M(q_s)^2 \frac{M(Q_AQ_C,q_s)M(Q_BQ_C,q_s)}{M(Q_A,q_s)M(Q_B,q_s)M(Q_C,q_s)M(Q_AQ_BQ_C,q_s)}.    \label{Ztop-tripleP1}
\ee
All non-zero Gopakumar-Vafa invariants for this geometry are of genus 0, and in class $\beta$ they read
$$
N_{\beta=0} = -4,\quad N_{\beta=\pm A} = N_{\beta=\pm B} = N_{\beta=\pm C} = N_{\beta=\pm (A + B + C)} = 1,
$$
$$
N_{\beta=\pm (B + C)} = N_{\beta=\pm (A + C)} = -1.   
$$

Let us specialize now the general structures from section \ref{sec-results} to the present case. 
We introduce four colors $q_g$, for $g\in\{b,c,a,0\}$ (instead of $q_1,q_2,q_3,q_0$ used earlier). We also often use
$$
\widehat{Q} = \widehat{Q}_a \widehat{Q}_b \widehat{Q}_c \widehat{Q}_0, \qquad \qquad q = q_0 q_a q_b q_c, 
$$
and to check some results against the ones for the resolved conifold also consider specialization
\be
q_0 \equiv q_c,\qquad \qquad q_1 \equiv q_a \equiv q_b.   \label{spec}
\ee

\begin{figure}[htb]
\begin{center}
\includegraphics[width=\textwidth]{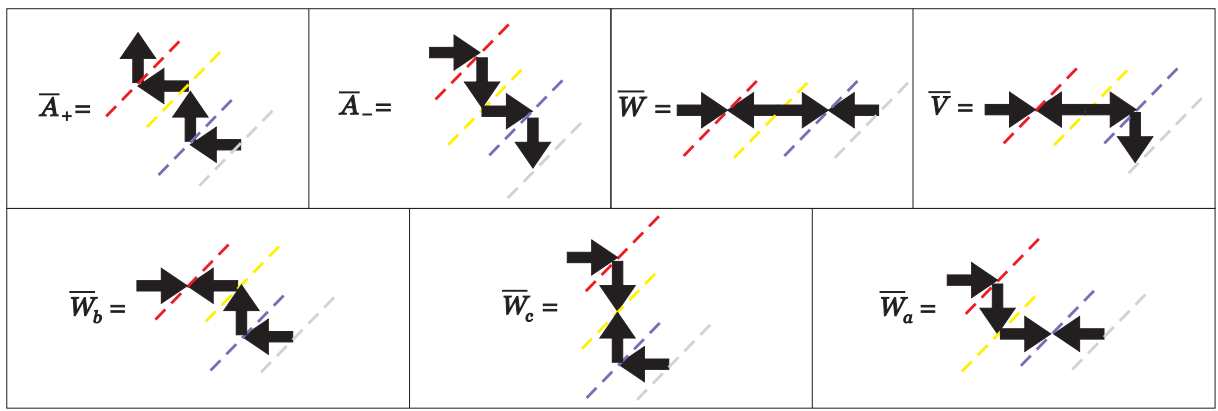} 
\begin{quote}
\caption{\emph{Building block of triple-$\mathbb{P}^1$ crystals. Operators $\overline{A}_{\pm}$ and $\overline{W}_{b}, \overline{W}_{c}, \overline{W}_{a}$ are the standard ones introduced in section \ref{sec-results}. Operators $W$ and $V$ are new ones, which change values of $B$-fields through all $\mathbb{P}^1$'s simultaneously and implement extensions of the top row of the pyramid crystal similarly as in the conifold case.}} \label{fig-tripleP1arrows}
\end{quote}
\end{center}
\end{figure}

Operators (\ref{Aplus}) and (\ref{Aminus}) take the form\footnote{The operators $A_{\pm}$, as well as the computation of (\ref{Z1}), appeared in \cite{YoungBryan} in the context of the closed topological vertex geometry. We discuss relations between both these geometries further in section \ref{ssec-ClosedVertex}.  \label{fn-Apm}}
$$
A_+(x) = \G_+(xq_0q_aq_bq_c) \G'_+(x q_0q_aq_c) \G_+(x q_0 q_a) \G'_+(xq_0),
$$
and 
$$
A_-(x) = \G_-(x) \G'_-(x q_b) \G_-(x q_b q_c) \G'_-(xq_aq_bq_c),
$$
and the commutation relation (\ref{A-commute}) specialize in this case to
\be
A_+(x) A_-(y) = A_-(y) A_+(x)  \times    \label{commute}
\ee
$$
\times \frac{(1+xy q q_b)(1+xy \frac{q}{q_b})(1+xy q q_c)(1+xy \frac{q}{q_c})(1+xy q q_a)(1+xy \frac{q}{q_a})(1+xy q_0)(1+xy \frac{q^2} {q_0})}{(1-xyq)^4 (1-xy q q_bq_c)(1-xy \frac{q}{q_bq_c}) (1-xy q q_aq_c)(1-xy \frac{q}{q_aq_c})}.  
$$
The ground states (\ref{Omega-plus}) and (\ref{Omega-minus}) are
\bea
|\Omega_-\rangle & = &  A_-(1) A_-(q) A_-(q^2)\ldots |0\rangle,  \\
\langle \Omega_+ | & = & \langle 0| \ldots A_+(q^2) A_+(q) A_+(1).
\eea
Operators $\overline{A}_{\pm}$, as well as $\overline{W}_{b}, \overline{W}_{c}, \overline{W}_{a}$, are translated in figure \ref{fig-tripleP1arrows} into the arrow form relevant to the construction of the crystals. We note that in the non-commutative Donaldson-Thomas chamber for $R>0$, the crystal has the same shape as the pyramid crystal for the conifold with one stone on top, however now it has four colors, see figure \ref{fig-triple-P1} (right). Its partition function in our framework can be found as (notation (\ref{MacMahons}) is used)
\be
Z^{triple}_1 = \langle \Omega_+ | \Omega_- \rangle = M(1,q)^4 \frac{\widetilde{M}(q_bq_c,q) \widetilde{M}(q_aq_c,q)}{\widetilde{M}(-q_a,q) \widetilde{M}(-q_b,q) \widetilde{M}(-q_c,q) \widetilde{M}(-q_aq_bq_c,q)}.  \label{Z1}
\ee
This follows from general results in section \ref{sec-results}, and of course can also be found directly$^{\textrm{\ref{fn-Apm}}}$ using commutation relations (\ref{commute}). Comparing with (\ref{Ztop-tripleP1}), we indeed find that
$$
Z^{triple}_1(q_a,q_b,q_c) = \cZ^{triple-\mathbb{P}^1}_{top}(Q_A,Q_B,Q_C)  \cZ^{triple-\mathbb{P}^1}_{top}(Q_A^{-1},Q_B^{-1},Q_C^{-1})
$$
under the following identification of parameters
$$
q_s = q,\qquad Q_A = -q_a,\qquad Q_B = -q_b,\qquad Q_C = -q_c.   
$$

As usual the extreme chamber with $R<0$ is represented by the empty pyramid with the generating function
$$
\widetilde{Z}^{triple}_1 = \langle 0 | 0 \rangle = 1.
$$

We introduce now new types of crystals which extend our prescription from section \ref{sec-results}, and associated to them new wall-crossing operators. As we already saw, in the non-commutative Donaldson-Thomas chamber the crystal with generating function (\ref{Z1}) consists of pyramid partitions of the same type as in the conifold, however with four colors. It is reasonable to conjecture that analogous infinite or finite four-colored pyramids, albeit (similarly as in the conifold case) with a string of arbitrary number of stones in the top row, should also provide BPS generating functions in certain chambers. Below we find generating functions of such crystals and show that they indeed reproduce BPS generating functions in chambers with particular choices of $B$-field through all three $\mathbb{P}^1$'s upon a simple change of variables.

\begin{figure}[htb]
\begin{center}
\includegraphics[width=0.9\textwidth]{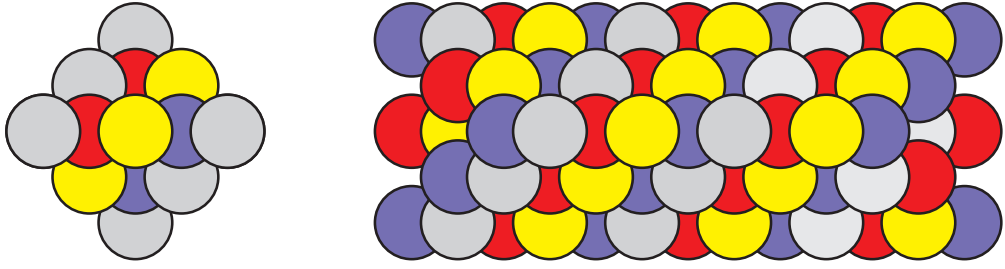} 
\begin{quote}
\caption{\emph{Four-colored pyramids: finite one with three stones in the top row, with generating function $\widetilde{Z}^{triple}_{4} = \langle 0 | \overline{W}^{2} |0\rangle$ (left), and infinite one with four stones in the top row and generating function $Z^{triple}_{4}  =  \langle \Omega_+ | \overline{W} \overline{V} |\Omega_- \rangle$ (right). Identifying variables as $q_a=q_b=q_1$ and $q_c=q_0$ leads to the conifold case.}} \label{fig-pyramids-inf}
\end{quote}
\end{center}
\end{figure}

It is clear from the figure \ref{fig-tripleP1arrows}, that neither the wall-crossing operators $\overline{W}_{b}, \overline{W}_{c}, \overline{W}_{a}$ which arise from our prescription for the triple-$\mathbb{P}^1$ geometry, nor their primed counterparts considered so far, could be used to construct pyramid of arbitrary length, such as shown in figure \ref{fig-pyramids-inf}. Therefore we have to introduce new types of operators. To preserve periodic four-colored pattern they should involve four vertex operators interlaced with $\widehat{Q}_g$ operators, and their pattern should follow from the assignment presented in figure \ref{fig-arrows}. Because there are four colors, the cases with even or odd stones in the top row are different. An inspection of figure \ref{fig-arrows} leads us to define the following operators
\bea
\overline{V} & = & \G_-(1) Q_b \G'_+(1) Q_c \G_-(1) Q_a \G'_-(1) Q_0, \nonumber \\   
\overline{W} & = & \G_-(1) Q_b \G'_+(1) Q_c \G_-(1) Q_a \G'_+(1) Q_0.                
\eea
Their translation to the arrow structure is shown in figure \ref{fig-tripleP1arrows} (first row on the right). 


Insertion of these operators extends the top row respectively by one or two stones. For arbitrary number of stones in the top row we should insert several operators $\overline{W}$ and possibly also a single $\overline{V}$. Let $Z^{triple}_n$ denote a generating function for a four-colored infinite pyramid with $n$ stones in the top row, and $\widetilde{Z}^{triple}_n$ a generating function for four-colored finite pyramid with $n-1$ stones in the top row. Therefore the generating functions in the infinite case read
\bea
Z^{triple}_{2n} & = & \langle \Omega_+ | \overline{W}^{n-1} \overline{V} |\Omega_- \rangle, \nonumber \\
Z^{triple}_{2n+1} & = & \langle \Omega_+ | \overline{W}^{n}|\Omega_- \rangle.
\eea
In particular $Z^{triple}_2 = \langle \Omega_+  | \overline{V} |\Omega_- \rangle$ and $Z^{triple}_3 = \langle \Omega_+  | \overline{W} |\Omega_+ \rangle$.
For the finite case
\bea
\widetilde{Z}^{triple}_{2n} & = & \langle 0 | \overline{W}^{n} |0\rangle, \nonumber \\
\widetilde{Z}^{triple}_{2n+1} & = & \langle 0 | \overline{W}^{n}\, \G_-(1)|0\rangle.
\eea

We compute now these partition functions explicitly, focusing first on the infinite case. To start with we determine ratios $Z^{triple}_{n+1}/Z^{triple}_n$. In fact there are two possibilities, depending on whether $n$ is even or odd. 
We note that
\bea
Z^{triple}_{2n+1} & = & \langle \Omega_+ | A_2^{n-1} | \G_-(1) {\bf \G'_+(\frac{1}{q_b})} \G_-(q_bq_c) {\bf \G'_+(\frac{1}{q_aq_bq_c})} | \widehat{Q}|\Omega_-\rangle, \nonumber \\
Z^{triple}_{2n} & = & \langle \Omega_+ | A_2^{n-1} | \G_-(1) {\bf \G'_+(\frac{1}{q_b})} \G_-(q_bq_c) {\bf \G'_-(q_aq_bq_c)} | \widehat{Q}|\Omega_-\rangle, \nonumber \\
Z^{triple}_{2n-1} & = & \langle \Omega_+ | A_2^{n-1} | \G_-(1) {\bf \G'_-(q_b)} \G_-(q_bq_c) {\bf \G'_-(q_aq_bq_c)} | \widehat{Q}|\Omega_-\rangle. \nonumber 
\eea
We see that the only difference between these quantities is in the form of $\G'_{\pm}$ operators (and their arguments) written in bold. So if commute ${\bf \G'_+}$ to the right and ${\bf \G'_-}$ to the left, the remaining correlators will be the same. In this way we find
\be
a_n := \frac{Z^{triple}_{2n+1}}{Z^{triple}_{2n}} =  (1-q_a q_c)\, \frac{\prod_{i=1}^{\infty} (1 + q^{i-1}q_0)(1+\frac{q^i}{q_a})}
{\prod_{i=n}^{\infty} (1 + \frac{q^{i+1}}{q_0})(1+q^i q_a)} \, \prod_{i=1}^{\infty} \frac{1-q^i q_a q_c}{1-\frac{q^i}{q_a q_c}}, \label{ratioa}
\ee
as well as
\be
b_n := \frac{Z^{triple}_{2n}}{Z^{triple}_{2n-1}} = \frac{1+q_c}{1-q_a q_c}\, \frac{\prod_{i=1}^{\infty} (1 + \frac{q^i}{q_b})(1+q^i q_c)}
{\prod_{i=n}^{\infty} (1 + q^i q_b)(1+\frac{q^i}{q_c})} \, \prod_{i=1}^{\infty} \frac{1-\frac{q^i}{q_a q_c}}{1-q^i q_a q_c}. \label{ratiob}
\ee
From these ratios and the value of $Z^{triple}_1$ given in (\ref{Z1}) we can reproduce explicit formulas for all $Z^{triple}_n$. We find
\bea
Z^{triple}_{2n+1} & = & Z^{triple}_1 \prod_{i=1}^n (a_i b_i) =      \label{Z2n1} \\
& = & M(1,q)^4 \widetilde{M}(q_a q_c,q)) \widetilde{M}(q_b q_c,q)) (1+q_c)^n  \times \nonumber \\
& & \times \prod_{i=1}^{\infty} \big(1+\frac{q^i}{q_a}\big)^{n+i}\big(1+\frac{q^i}{q_b}\big)^{n+i}\big(1+q_c q^i)^{n+i}\big(1+\frac{q^i}{q_a q_b q_c}\big)^{n+i}      \times \nonumber \\
& & \times \prod_{i=n+1}^{\infty} \big(1+ q_a q^i\big)^{i-n}\big(1+q_b q^i\big)^{i-n}\big(1+\frac{q^i}{q_c}\big)^{i-n}\big(1+q_a q_b q_c q^i\big)^{i-n},  \nonumber
\eea
as well as
\bea
Z^{triple}_{2n} & = & Z^{triple}_1 b_n \prod_{i=1}^{n-1} (a_i b_i) =      \label{Z2n} \\
& = & M(1,q)^4 \widetilde{M}(q_a q_c,q)) \widetilde{M}(q_b q_c,q)) \frac{(1+q_c)^n}{1-q_a q_c} \prod_{i=1}^{\infty}\frac{1-\frac{q^i}{q_a q_c}}{1- q_a q_c q^i} \times \nonumber \\
& & \times \prod_{i=1}^{\infty} \big(1+\frac{q^i}{q_a}\big)^{n+i-1}\big(1+\frac{q^i}{q_b}\big)^{n+i}\big(1+q_c q^i)^{n+i}\big(1+\frac{q^i}{q_a q_b q_c}\big)^{n+i-1}      \times \nonumber \\
& & \times \prod_{i=n}^{\infty} \big(1+ q_a q^i\big)^{i-n+1}\big(1+q_b q^i\big)^{i-n}\big(1+\frac{q^i}{q_c}\big)^{i-n}\big(1+q_a q_b q_c q^i\big)^{i-n+1}.  \nonumber
\eea
We checked that upon specialization (\ref{spec}) both formulas above reduce to the generating functions of two-colored pyramid partitions with the same number of stones in the top row (\ref{Zn}), as they indeed should.

In the finite case we have
\bea
\widetilde{Z}^{triple}_{2n+1} & = & \langle 0 | A_2^{n-1} | \G_-(1) \G'_+(\frac{1}{q_b}) \G_-(q_bq_c) \G'_+(\frac{1}{q_aq_bq_c}) \G_-(q) | 0\rangle, \nonumber \\
\widetilde{Z}^{triple}_{2n} & = & \langle 0 | A_2^{n-1} | \G_-(1) \G'_+(\frac{1}{q_b}) \G_-(q_bq_c) | 0\rangle, \nonumber \\
\widetilde{Z}^{triple}_{2n-1} & = & \langle 0 | A_2^{n-1} | \G_-(1) | 0\rangle. \nonumber 
\eea
Commuting these expressions we find
\be
\tilde{a}_n := \frac{\widetilde{Z}^{triple}_{2n+1}}{\widetilde{Z}^{triple}_{2n}} =  \prod_{i=0}^{n-1} (1 + q^{i}q_0)(1+\frac{q^{i+1}}{q_b}), \label{ratioatilde}
\ee
as well as
\be
\tilde{b}_n := \frac{\widetilde{Z}^{triple}_{2n}}{\widetilde{Z}^{triple}_{2n-1}} = \frac{q_a}{1+q_a} \prod_{i=0}^{n-1} (1 + \frac{q^{i}}{q_a})(1+q^{i} q_c). \label{ratiobtilde}
\ee
Together with $\widetilde{Z}^{triple}_1=1$ it now follows that
\bea
\widetilde{Z}^{triple}_{2n+1} & = & \prod_{i=1}^n (\tilde{a}_i \tilde{b}_i) =      \label{Z2n1tilde} \\
& = & \frac{q_a^n}{(1+q_a)^n}  
\prod_{i=1}^{n} \big(1+\frac{q^{n-i}}{q_a}\big)^{i}\big(1+\frac{q^{n-i+1}}{q_b}\big)^{i}\big(1+q_c q^{n-i})^{i}\big(1+q^{n-i} q_0\big)^{i} ,   \nonumber
\eea
as well as
\bea
\widetilde{Z}^{triple}_{2n} & = &  \tilde{b}_n \prod_{i=1}^{n-1} (\tilde{a}_i \tilde{b}_i) =      \label{Z2ntilde} \\
& = & \frac{q_a^n}{(1+q_a)^n}  
\prod_{i=1}^{n} \big(1+\frac{q^{n-i}}{q_a}\big)^{i}\big(1+\frac{q^{n-i+1}}{q_b}\big)^{i-1}\big(1+q_c q^{n-i})^{i}\big(1+q^{n-i} q_0\big)^{i-1} .   \nonumber
\eea
Upon specialization (\ref{spec}) both these expressions reduce to the generating functions of two-colored pyramid partitions with the same number of stones in the top row (\ref{ZZn}), as they should.

To sum up, we have found the generating functions of four-colored finite and infinite pyramid partitions, with arbitrary (even or odd) number of stones in the top row. The task that remains is to show that there exist such values of moduli $R$ and $B$, and such identification between crystal and string parameters, so that the structure of these generating functions is consistent with the structure (\ref{cZ-chamber}) encoded in the topological string partition function, together with the condition on central charges (\ref{Zpositive}). Below we show that these conditions are indeed met, and therefore the above crystal generating functions do provide the correct BPS counting functions. We consider finite/infinite and even/odd cases separately.


\subsubsection*{$Z^{triple}_{2n+1}$: infinite pyramid, odd chambers}

To start with we consider the generating function for an infinite pyramid with odd number of boxes $2n+1$ in the top row (\ref{Z2n1}). Comparison of the form of products involving $(1+q_a^{\pm 1}q^i)$, $(1+q_b^{\pm 1}q^i)$ and $(1+q_c^{\pm 1}q^i)$ with the chambers in the resolved conifold suggests the following the identification of the string coupling $q_s=e^{-g_s}$ and K\"ahler parameters:
\be
q_s = q,\qquad Q_A = -q_a q^n,\qquad Q_B = -q_b q^n,\qquad Q_C = -q_c q^{-n},    \label{redefine-inf2n1}
\ee
as well as $R>0$ as was the case for infinite pyramids in the conifold case as well.
In particular, under this identification $\prod_{i=n+1}^{\infty}(1+q_aq_bq_cq^i)^{i-n}=\prod_{i=1}^{\infty}(1-Q_AQ_BQ_Cq_s^i)^i$, $\widetilde{M}(q_aq_c,q)=\widetilde{M}(Q_AQ_C,q_s)$, $\widetilde{M}(q_bq_c,q)=\widetilde{M}(Q_BQ_C,q_s)$, $(1+q_c)^n\prod_{i=1}^{\infty}(1+q_cq^i)^{n+i} = \prod_{j=n}^{\infty}(1-Q_Cq_s^j)^j$, etc. Now the form of products involving single $Q_A, Q_B$ and $Q_C$ leads to the following identification of moduli
\be
R>0,\qquad n<B_A<n+1,\qquad n<B_B<n+1,\qquad -n<B_C<-n+1,\qquad \textrm{for}\ n\geq 1.  \label{impose-inf2n1}
\ee
This is consistent with the form of products involving $Q_AQ_C$, $Q_BQ_C$ and $Q_AQ_BQ_C$, however they impose respectively the following additional constraints: 
\be
0<B_A+B_C<1, \qquad 0<B_B+B_C<1, \qquad n<B_A+B_B+B_C<n+1,   \label{impose-inf2n1-bis}
\ee
which implies that in fact $2n<B_A+B_B<2n+1$.
Finally, the fact that $R>0$ implies that all factors $M(1,q)$ should indeed be present. To sum up, under identifications (\ref{redefine-inf2n1}), the crystal model for infinite pyramid with $2n+1$ boxes in the top row leads to the generating function
\bea
\cZ^{triple}_{2n+1} & = & M(1,q_s)^4 \widetilde{M}(Q_AQ_C,q_s) \widetilde{M}(Q_BQ_C,q_s) \times \nonumber \\
& & \times \prod_{i=1}^{\infty} \big(1-Q_A q_s^i\big)^{i}\big(1-Q_Bq_s^i\big)^{i}\big(1-Q_C^{-1} q_s^i)^{i}\big(1-Q_AQ_BQ_Cq_s^i\big)^{i}      \times \nonumber \\
& & \times \prod_{i=n+1}^{\infty} \big(1-Q_A^{-1} q_s^i\big)^{i}\big(1-Q_B^{-1}q_s^i\big)^{i} \big(1-Q_C q_s^{i-1}\big)^{i-1}\big(1-Q_A^{-1}Q_B^{-1}Q_C^{-1}q_s^i\big)^{i}, \nonumber
\eea
which is indeed the consistent generating function of D6-D2-D0 bounds states in the triple-$\mathbb{P}^1$ geometry, in chambers specified by (\ref{impose-inf2n1}) and (\ref{impose-inf2n1-bis}).


\subsubsection*{$Z^{triple}_{2n}$: infinite pyramid, even chambers}

Next we consider the generating function for an infinite pyramid with even number of boxes $2n$ in the top row (\ref{Z2n}). Now the identification of the string coupling $q_s$ and K\"ahler parameters reads:
\be
q_s = q,\qquad Q_A = -q_a q^{n-1},\qquad Q_B = -q_b q^n,\qquad Q_C = -q_c q^{-n}.    \label{redefine-inf2n}
\ee
The form of products involving $Q_A, Q_B$ and $Q_C$ leads to the following identification of moduli
\be
R>0,\qquad n-1<B_A<n,\qquad n<B_B<n+1,\qquad -n<B_C<-n+1,\qquad \textrm{for}\ n\geq 1.  \label{impose-inf2n}
\ee
together with additional constraints: 
\be
0<B_A+B_C<1, \qquad 0<B_B+B_C<1, \qquad n-1<B_A+B_B+B_C<n.   \label{impose-inf2n-bis}
\ee
Again $R>0$ implies that all factors $M(1,q_s)$ are present. To sum up, in this case the crystal model leads to the generating function
\bea
\cZ^{triple}_{2n} & = & M(1,q_s)^4 \widetilde{M}(Q_AQ_C,q_s) \widetilde{M}(Q_BQ_C,q_s) \times \nonumber \\
& & \times \prod_{i=1}^{\infty} \big(1-Q_A q_s^i\big)^{i}\big(1-Q_Bq_s^i\big)^{i}\big(1-Q_C^{-1} q_s^i)^{i}\big(1-Q_AQ_BQ_Cq_s^i\big)^{i}      \times \nonumber \\
& & \times \prod_{i=n}^{\infty} \big(1-Q_A^{-1} q_s^i\big)^{i}\big(1-Q_B^{-1}q_s^{i+1}\big)^{i+1} \big(1-Q_C q_s^{i}\big)^{i}\big(1-Q_A^{-1}Q_B^{-1}Q_C^{-1}q_s^i\big)^{i}, \nonumber
\eea
which is indeed the generating function of D6-D2-D0 bounds states in the triple-$\mathbb{P}^1$ geometry, in chambers specified by (\ref{impose-inf2n}) and (\ref{impose-inf2n-bis}).


\subsubsection*{$\widetilde{Z}^{triple}_{2n+1}$: finite pyramid, odd chambers}

Now we consider the generating function for a finite pyramid in odd odd chambers labeled by $2n+1$ (i.e. the pyramid with $2n$ stones in the top row), given by (\ref{Z2n1tilde}). The identification of the string coupling $q_s$ and K\"ahler parameters reads:
\be
q_s = \frac{1}{q},\qquad Q_A = -q_a q^{-n},\qquad Q_B = -q_b q^{-n-1},\qquad Q_C = -q_c q^{n},    \label{redefine-fin2n1}
\ee
and now $R<0$. The form of products involving $Q_A, Q_B$ and $Q_C$ leads to the following identification of moduli
\be
R<0,\qquad n-1<B_A<n,\qquad n<B_B<n+1,\qquad -n-1<B_C<-n,\qquad \textrm{for}\ n\geq 1.  \label{impose-fin2n1}
\ee
together with additional constraint: 
\be
-1<B_A+B_C<0,\quad -1<B_B+B_C<0  ,\qquad  n<B_A+B_B+B_C<n+1.   \label{impose-fin2n1-bis}
\ee
In particular $R<0$ implies that factors of $M(1,q_s)$ should indeed be absent in this case. 

Altogether, in this case the crystal model leads to the generating function
\be
\widetilde{\cZ}^{triple}_{2n+1} = \prod_{i=1}^{n} \big(1-Q_A^{-1} q_s^{i-1}\big)^{i-1}\big(1-Q_B^{-1}q_s^i\big)^{i}\big(1-Q_C q_s^i)^{i}\big(1-Q_A^{-1}Q_B^{-1}Q_C^{-1}q_s^i\big)^{i}.      \nonumber 
\ee
This is indeed the generating function of D6-D2-D0 bounds states in the triple-$\mathbb{P}^1$ geometry, in chambers specified above.


\subsubsection*{$\widetilde{Z}^{triple}_{2n}$: finite pyramid, even chambers}

Finally we consider the generating function for a finite pyramid in even chambers labeled by $2n$ (i.e. with $2n-1$ stones in the top row), given by (\ref{Z2ntilde}). The identification of the string coupling $q_s$ and K\"ahler parameters reads:
\be
q_s = \frac{1}{q},\qquad Q_A = -q_a q^{-n},\qquad Q_B = -q_b q^{-n},\qquad Q_C = -q_c q^{n}.    \label{redefine-fin2n}
\ee
The form of products involving $Q_A, Q_B$ and $Q_C$ leads to the following identification of moduli
\be
R<0,\qquad n-1<B_A<n,\qquad n-1<B_B<n,\qquad -n-1<B_C<-n,\qquad \textrm{for}\ n\geq 1.  \label{impose-fin2n}
\ee
together with additional constraint: 
\be
-1<B_A+B_C<0,\quad -1<B_B+B_C<0, \qquad  n-1<B_A+B_B+B_C<n.   \label{impose-fin2n-bis}
\ee
In particular $R<0$ implies that factors of $M(1,q_s)$ should indeed be absent in this case. 

Therefore, in this case the crystal model leads to the generating function
\be
\widetilde{Z}^{triple}_{2n} =   (1-Q_C q_s) \prod_{i=1}^{n-1} \big(1-Q_A^{-1} q_s^{i}\big)^{i}\big(1-Q_B^{-1}q_s^i\big)^{i}\big(1-Q_C q_s^{i+1})^{i+1}\big(1-Q_A^{-1}Q_B^{-1}Q_C^{-1}q_s^i\big)^{i}.      \nonumber 
\ee
This is indeed the generating function of D6-D2-D0 bounds states in the triple-$\mathbb{P}^1$ geometry, in chambers specified above.




\subsection{Closed topological vertex}   \label{ssec-ClosedVertex}

\begin{figure}[htb]
\begin{center}
\includegraphics[width=0.4\textwidth]{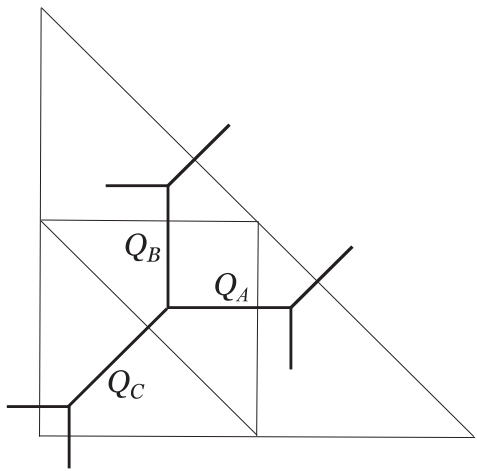} 
\begin{quote}
\caption{\emph{Toric diagram for the closed topological vertex geometry.}} \label{fig-closedvertex}
\end{quote}
\end{center}
\end{figure}

In this section we discuss wall-crossing for the closed topological vertex geometry, whose toric diagram is shown in figure \ref{fig-closedvertex}. We will denote various quantities associated to it by the label $\cC$. This geometry is also the symmetric resolution of $\mathbb{C}^3 / \mathbb{Z}_2\times \mathbb{Z}_2$ orbifold. This is an example of the geometry which does not arise from the triangulation of the strip. Nonetheless it does not contain compact four-cycles, therefore the physical arguments reviewed in section \ref{ssec-wallcross} apply in this case as well. 

The closed topological vertex shares important similarities with the triple-$\mathbb{P}^1$ geometry analyzed in the previous section. 
In this case let us also denote by $A,B,C$ the three independent $\mathbb{P}^1$'s, and by $T_A, T_B, T_C$ the corresponding K{\"a}hler parameters, and let
$$
Q_A = e^{-T_A}, \qquad  Q_B = e^{-T_B}, \qquad  Q_C = e^{-T_C}.
$$
In particular the topological string partition function for the closed vertex differs from the one of the triple-$\mathbb{P}^1$ (\ref{Ztop-tripleP1}) just by one factor of MacMahon function
\bea
\cZ^{\cC}_{top}(Q_A,Q_B,Q_C) & = & M(q_s)^2 \frac{M(Q_AQ_B,q_s)M(Q_AQ_C,q_s)M(Q_BQ_C,q_s)}{M(Q_A,q_s)M(Q_B,q_s)M(Q_C,q_s)M(Q_AQ_BQ_C,q_s)} = \nonumber \\
& = &  M(Q_AQ_B,q_s) \cdot \cZ^{triple}_{top}(Q_A,Q_B,Q_C).     \label{ZC-Ztriple-tops}
\eea
From this we find that all non-zero Gopakumar-Vafa invariants for $\cC$ are of genus 0, and in class $\beta$ they read
$$
N_{\beta=0} = -4,\quad N_{\beta=\pm A} = N_{\beta=\pm B} = N_{\beta=\pm C} = N_{\beta=\pm (A + B + C)} = 1,  
$$
$$
N_{\beta=\pm (A + B)} =  N_{\beta=\pm (B + C)} = N_{\beta=\pm (A + C)} = -1.    \nonumber 
$$

The close relation between partition functions of $\cC$ and triple-$\mathbb{P}^1$ suggests the existence of similar underlying crystal models in both cases. There is a natural crystal model for the non-commutative Donaldson-Thomas chamber for $\cC$, discussed in \cite{YoungBryan}, which consists of four-colored plane partitions colored according to the action of $\mathbb{Z}_2\times \mathbb{Z}_2$ group. Let us denote the generating function of these partitions by $Z^{\cC}$. In \cite{YoungBryan} it was shown that $Z^{\cC}$ indeed reproduces the non-commutative Donaldson-Thomas invariants for $\cC$. Moreover, it was also shown that it is closely related to four-colored partitions (with one stone on top) which we associated to triple-$\mathbb{P}^1$ in the previous section. In particular the generating functions of these two crystal models differ by a factor $\widetilde{M}(q_a q_b,q)$
\be
Z^{\cC}  = \widetilde{M}(q_a q_b,q) Z^{triple}_1,    \label{ZC-Ztriple}
\ee
where $Z^{triple}_1$ is given by the formula (\ref{Z1}). This relation is of course consistent with (\ref{ZC-Ztriple-tops}).

Our aim now is to extend the relation between the non-commutative Donaldson-Thomas invariants for these two geometries beyond the extreme chamber. We focus first on the chambers for triple-$\mathbb{P}^1$ associated to extended pyramid partitions from the previous section. We postulate that a generating function of each such partition can be identified with a generating function of BPS invariants of $\cC$ in certain chamber, up to some overall factor of the form
\be
\mu_n(Q_AQ_B,q_s).   \label{mu-n}
\ee
This overall factor should reduce to $\widetilde{M}(q_a q_b)$ to reproduce (\ref{ZC-Ztriple}) in the extreme chamber. 
Moreover, we postulate that a similar relations hold for finite pyramids, and BPS generating functions for $\cC$ and triple-$\mathbb{P}^1$ in chambers corresponding to finite four-colored pyramids are identified up to a factor of a form
\be
\widetilde{\mu}_n(Q_AQ_B,q_s), \label{mu-tilde-n}
\ee
which reduces to 1 in the extreme chamber with a pure D6-brane.

We will apply now the physical arguments of section \ref{ssec-wallcross} to prove that these postulates are true.
The closed vertex BPS generating functions $\cZ^{\cC}(Q_A,Q_B,Q_C; B,R)$ and $\widetilde{\cZ}^{\cC}(Q_A,Q_B,Q_C; B,R)$ associated respectively to chambers corresponding to infinite and finite pyramids depend on the values of four moduli: the radius $R$ and $B$-fields through three $\mathbb{P}^1$'s which we denote by $B_A,B_B,B_C$. We prove that these postulates are true by finding the explicit form, for each such pyramid, of:
\begin{itemize}
\item an identification between crystal $q_0,q_a,q_b,q_c$ and string $q_s,Q_A,Q_B,Q_C$ parameters,
\item values of moduli $R, B_A, B_B, B_C$,
\item correction factors $\mu_n(q_a q_b,q)$ and $\widetilde{\mu}_n(q_a q_b,q)$ for infinite and finite pyramids,
\end{itemize}
such that 
\bea
Z^{triple}_{n}(q_0,q_a,q_b,q_c) \, \mu_n(q_a q_b,q) = \cZ^{\cC}(Q_A,Q_B,Q_C; B,R),  \label{guess1}  \\
\widetilde{Z}^{triple}_{n}(q_0,q_a,q_b,q_c) \, \widetilde{\mu}_n(q_a q_b,q) = \widetilde{\cZ}^{\cC}(Q_A,Q_B,Q_C; B,R), \label{guess2}   
\eea
respectively for infinite and finite pyramids. The identification of crystal and string parameters, as well as values of stringy moduli, must be completely specified by $n$ which encodes the length of a given pyramid.

In fact, as a consequence of the relation between partition functions (\ref{ZC-Ztriple-tops}) we realized the above postulates to some extent already in the previous section, by matching four-colored pyramid generating functions to BPS counting functions for triple-$\mathbb{P}^1$. All we have to show now is that there exist appropriate $\mu_n(Q_AQ_B,q_s)$ and $\widetilde{\mu}_n(Q_AQ_B,q_s)$, consistent with the values of moduli, which is a non-trivial condition. Nonetheless this is indeed the case, as we show below for each case of infinite/finite and even/odd chambers. 

While the identification of wall-crossing chambers completed below is satisfying from the physical point of view, there still remain several interesting questions. 

Firstly, one might wonder what are the crystals whose generating functions are given by the left hand side of equations (\ref{guess1}) and (\ref{guess2}). As reviewed above, in the extreme chamber for infinite pyramids such a crystal in given by $\mathbb{Z}_2 \times \mathbb{Z}_2$-four-colored plane partitions. We conjecture that for all other chambers discussed above, the corresponding crystal model is given in terms of similar four-colored plane partitions, which fill a container that develops appropriate number of corners, similarly as in figure \ref{fig-OrbiC3}.

Secondly, it would be interesting to find whether there exist, and if so that what are crystal models associated to other chambers of the closed topological vertex.


\subsubsection*{$\cZ^{\cC}_{2n+1}$: infinite pyramid, odd chambers}

We consider first the chambers corresponding to infinite pyramids with $2n+1$ stones in the top row with generating functions (\ref{Z2n1}). We postulate that the relations (\ref{redefine-inf2n1}), (\ref{impose-inf2n1}) and (\ref{impose-inf2n1-bis}) found for the triple-$\mathbb{P}^1$ hold also for the closed vertex:
$$
q_s = q,\qquad Q_A = -q_a q^n,\qquad Q_B = -q_b q^n,\qquad Q_C = -q_c q^{-n},
$$
$$
R>0,\qquad n<B_A<n+1,\qquad n<B_B<n+1,\qquad -n<B_C<-n+1,\qquad \textrm{for}\ n\geq 1,
$$
$$
0<B_A+B_C<1, \qquad 0<B_B+B_C<1, \qquad n<B_A+B_B+B_C<n+1. 
$$
In particular this implies that $2n<B_A+B_B<2n+1$, and therefore the factor (\ref{mu-n}) 
can be chosen consistently as
$$
\mu_{2n+1}(Q_AQ_B,q_s) = \prod_{i=1}^{\infty} \frac{1}{(1-Q_AQ_Bq_s^i)^i} \, \prod_{i=2n+1}^{\infty} \frac{1}{(1-Q_A^{-1}Q_B^{-1}q_s^i)^i}.
$$
Finally, the fact that $R>0$ implies that all factors $M(1,q)$ should indeed be present. To sum up, under the above identifications, 
the crystal model for infinite pyramid with $2n+1$ boxes in the top row leads to the BPS generating function
\bea
\cZ^{\cC}_{2n+1} & = & M(1,q_s)^4 \widetilde{M}(Q_AQ_C,q_s) \widetilde{M}(Q_BQ_C,q_s) \mu_{2n+1}(Q_AQ_B,q_s) \times \nonumber \\
& & \times \prod_{i=1}^{\infty} \big(1-Q_A q_s^i\big)^{i}\big(1-Q_Bq_s^i\big)^{i}\big(1-Q_C^{-1} q_s^i)^{i}\big(1-Q_AQ_BQ_Cq_s^i\big)^{i}      \times \nonumber \\
& & \times \prod_{i=n+1}^{\infty} \big(1-Q_A^{-1} q_s^i\big)^{i}\big(1-Q_B^{-1}q_s^i\big)^{i} \big(1-Q_C q_s^{i-1}\big)^{i-1}\big(1-Q_A^{-1}Q_B^{-1}Q_C^{-1}q_s^i\big)^{i}, \nonumber
\eea
and this is indeed a consistent generating function of D6-D2-D0 bounds states for the closed topological vertex, in chambers specified above.


\subsubsection*{$\cZ^{\cC}_{2n}$: infinite pyramid, even chambers}

Next we consider the chambers related to infinite pyramids with $2n$ stones in the top row, with generating functions (\ref{Z2n}). 
Again we assume that the relations (\ref{redefine-inf2n}), (\ref{impose-inf2n}) and (\ref{impose-inf2n-bis}) found for the triple-$\mathbb{P}^1$ hold for the closed vertex:
$$
q_s = q,\qquad Q_A = -q_a q^{n-1},\qquad Q_B = -q_b q^n,\qquad Q_C = -q_c q^{-n},
$$
$$
R>0,\qquad n-1<B_A<n,\qquad n<B_B<n+1,\qquad -n<B_C<-n+1,\qquad \textrm{for}\ n\geq 1,
$$
$$
0<B_A+B_C<1, \qquad 0<B_B+B_C<1, \qquad n-1<B_A+B_B+B_C<n.
$$

These values of background fields also imply that the factor (\ref{mu-n}) can be chosen consistently as
$$
\mu_{2n}(Q_AQ_B,q_s) = \prod_{i=1}^{\infty} \frac{1}{(1-Q_AQ_Bq_s^i)^i} \, \prod_{i=2n}^{\infty} \frac{1}{(1-Q_A^{-1}Q_B^{-1}q_s^i)^i}.
$$
The fact that $R>0$ implies that all factors $M(1,q_s)$ are present. To sum up, in this case the crystal model leads to the BPS generating function
\bea
\cZ^{\cC}_{2n} & = & M(1,q_s)^4 \widetilde{M}(Q_AQ_C,q_s) \widetilde{M}(Q_BQ_C,q_s) \mu_{2n}(Q_AQ_B,q_s) \times \nonumber \\
& & \times \prod_{i=1}^{\infty} \big(1-Q_A q_s^i\big)^{i}\big(1-Q_Bq_s^i\big)^{i}\big(1-Q_C^{-1} q_s^i)^{i}\big(1-Q_AQ_BQ_Cq_s^i\big)^{i}      \times \nonumber \\
& & \times \prod_{i=n}^{\infty} \big(1-Q_A^{-1} q_s^i\big)^{i}\big(1-Q_B^{-1}q_s^{i+1}\big)^{i+1} \big(1-Q_C q_s^{i}\big)^{i}\big(1-Q_A^{-1}Q_B^{-1}Q_C^{-1}q_s^i\big)^{i},  \nonumber
\eea
which is a consistent generating function of D6-D2-D0 bounds states for the closed topological vertex, in chambers specified by (\ref{impose-inf2n}) and (\ref{impose-inf2n-bis}).


\subsubsection*{$\widetilde{\cZ}^{\cC}_{2n+1}$: finite pyramid, odd chambers}

We turn to consider the generating function for finite pyramids associated to odd $2n+1$ chambers, with generating functions (\ref{Z2n1tilde}). From the analysis of triple-$\mathbb{P}^1$ we know that (\ref{redefine-fin2n1}), (\ref{impose-fin2n1}) and (\ref{impose-fin2n1-bis}) should also hold:
$$
q_s = \frac{1}{q},\qquad Q_A = -q_a q^{-n},\qquad Q_B = -q_b q^{-n-1},\qquad Q_C = -q_c q^{n},
$$
$$
R<0,\qquad n-1<B_A<n,\qquad n<B_B<n+1,\qquad -n-1<B_C<-n,\qquad \textrm{for}\ n\geq 1,
$$
$$
-1<B_A+B_C<0,\quad -1<B_B+B_C<0  ,\qquad  n<B_A+B_B+B_C<n+1.
$$

Now $R<0$ implies that factors of $M(1,q_s)$ should be absent. The above values of background fields imply that the factor (\ref{mu-tilde-n}) should be chosen as
$$
\widetilde{\mu}_{2n+1}(Q_AQ_B,q_s) = \prod_{i=1}^{2n} \frac{1}{(1-Q_A^{-1}Q_B^{-1}q_s^i)^i}.
$$
Altogether, in this case the crystal model leads to the BPS generating function
\be
\widetilde{\cZ}^{\cC}_{2n+1} =  \widetilde{\mu}_{2n+1}(Q_AQ_B,q_s)  
\prod_{i=1}^{n} \big(1-Q_A^{-1} q_s^{i-1}\big)^{i-1}\big(1-Q_B^{-1}q_s^i\big)^{i}\big(1-Q_C q_s^i)^{i}\big(1-Q_A^{-1}Q_B^{-1}Q_C^{-1}q_s^i\big)^{i}.      \nonumber 
\ee
This is the consistent generating function of D6-D2-D0 bounds states in the closed topological vertex geometry, in chambers specified above.


\subsubsection*{$\widetilde{\cZ}^{\cC}_{2n}$: finite pyramid, even chambers}

Finally we consider the generating functions associated with finite pyramids and even $2n$ chambers, with generating functions (\ref{Z2ntilde}). The analysis of triple-$\mathbb{P}^1$ case implies (\ref{redefine-fin2n}), (\ref{impose-fin2n}) and (\ref{impose-fin2n-bis})
$$
q_s = \frac{1}{q},\qquad Q_A = -q_a q^{-n},\qquad Q_B = -q_b q^{-n},\qquad Q_C = -q_c q^{n},
$$
$$
R<0,\qquad n-1<B_A<n,\qquad n-1<B_B<n,\qquad -n-1<B_C<-n,\qquad \textrm{for}\ n\geq 1,
$$
$$
-1<B_A+B_C<0,\quad -1<B_B+B_C<0, \qquad  n-1<B_A+B_B+B_C<n.
$$

Now $R<0$ implies that factors of $M(1,q_s)$ should indeed be absent in this case. 
These values of background fields also imply that the factor (\ref{mu-tilde-n}) should take the form
$$
\widetilde{\mu}_{2n}(Q_AQ_B,q_s) = \prod_{i=1}^{2n-1} \frac{1}{(1-Q_A^{-1}Q_B^{-1}q_s^i)^i}.
$$
Altogether, in this case the crystal model leads to the generating function
\be
\widetilde{\cZ}^{\cC}_{2n} =  \widetilde{\mu}_{2n}(Q_AQ_B,q_s)  (1-Q_C q_s)
\prod_{i=1}^{n-1} \big(1-Q_A^{-1} q_s^{i}\big)^{i}\big(1-Q_B^{-1}q_s^i\big)^{i}\big(1-Q_C q_s^{i+1})^{i+1}\big(1-Q_A^{-1}Q_B^{-1}Q_C^{-1}q_s^i\big)^{i}.      \nonumber 
\ee
This is a consistent generating function of D6-D2-D0 bounds states for the closed topological vertex, in chambers specified above.


\section{Proofs}    \label{sec-proofs}

In this section we provide proofs of statements from section \ref{sec-results}, which relate BPS generating functions with fermionic correlators. 


\subsection{Quantization of geometry}    \label{ssec-ProofQuantize}

In this section we prove (\ref{Z-cZ}), which states that
$$
Z = \cZ, 
$$
under the identification of parameters (\ref{qQ}). Here  
$$
Z = \langle \Omega_+ | \Omega_- \rangle,
$$
is the overlap of the states $|\Omega_{\pm}\rangle$ which encode information about the classical geometry and $\cZ$ denotes the BPS partition function in the non-commutative Donaldson-Thomas chamber.

The states $| \Omega_{\pm} \rangle$ are defined in terms of $A_{\pm}$ as in (\ref{Omega-plus}) and (\ref{Omega-minus}). Using (\ref{Qcommute}) one can check that $A_{\pm}$ satisfy the commutation relation
\be
A_+(x) A_-(y) = C(x,y) A_-(y) A_+(x),     \label{A-commute}
\ee
with
$$
C(x,y) = \frac{1}{(1-xyq)^{N+1}} \prod_{1\leq i < j \leq N+1} \Big[\big(1- t_it_j\, xyq\, (q_i q_{i+1}\cdots q_{j-1}) \big)\big(1- t_it_j \frac{xyq}{q_i q_{i+1}\cdots q_{j-1}}  \big) \Big]^{-t_i t_j} .     
$$

The proof of (\ref{Z-cZ}) therefore amounts to repeated application of relations (\ref{A-commute}) in order to commute all $\G_+$ operators to the right of $\G_-$, similarly as in the case of MacMahon function and (\ref{C3-crystal}). We get
\begin{align}
& \langle \Omega_+ | \Omega_- \rangle = \langle 0 | \ldots A_+(q^2) A_+(q) A_+(1) | A_-(1) A_-(q) A_-(q^2) \ldots |0\rangle  = \prod_{r,s=0}^{\infty} C(q^r,q^s) =    \nonumber \\
= & \, M(1,q)^{N+1} \prod_{l=1}^{\infty} \prod_{1\leq i < j \leq N+1} \Big[\big(1- t_it_j\, q^l\, (q_i q_{i+1}\cdots q_{j-1}) \big)^l \big(1- t_it_j \frac{q^l}{q_i q_{i+1}\cdots q_{j-1}} \big)^l \Big]^{-t_i t_j}. \nonumber 
\end{align}
This expression indeed reproduces $\cZ=\cZ_{top}(Q_i) \cZ_{top}(Q^{-1}_i)$, with $\cZ_{top}(Q_i)$ given in (\ref{Ztop-strip}) and changing variables according to (\ref{qQ}). In particular, under the redefinition (\ref{qQ}) each factor of the infinite product becomes
$$
\big(1-(t_it_j)(t_it_{i+1}Q_i^{\pm 1})(t_{i+1}t_{i+2}Q_{i+1}^{\pm 1})\cdots(t_{j-1}t_{j}Q_{j-1}^{\pm 1}) q_s ^l\big)^{- t_i t_j l} = (1-(Q_i\cdots Q_{j-1})^{\pm 1} q_s ^l)^{-t_i t_j l}.
$$
Therefore one always gets an overall minus sign inside the bracket, as expected for $\cZ_{top}(Q_i^{\pm 1})$. On the other hand, the overall power is either $+l$ or $-l$, if $i$'th and $j$'th vertex are respectively of the opposite or of the same type. This means that a factor we consider appears respectively in numerator or denominator, in accordance with the "effective" local neighborhood of a chain of $\mathbb{P}^1$'s $Q_i\cdots Q_{j-1}$ being respectively $\mathcal{O}(-1)\oplus\mathcal{O}(-1)$ or $\mathcal{O}(-2)\oplus\mathcal{O}$. This proves (\ref{Z-cZ}).


\subsection{Wall-crossing operators}  \label{ssec-ProofWalls}

In this section we prove statements from section \ref{ssec-WallCross}, considering respectively all possible signs of $R$ and $B$-field. In all formulas below we represent operators $\overline{W}_p$, $\overline{W}'_p$ and $A_{\pm}$ as in (\ref{W-gamma}) and (\ref{Apm-gamma}), with the same fixed $p$.


\subsubsection*{Chambers with $R<0$, $B>0$}


Here we prove the equality (\ref{ZRnegBpos-Results}) and in consequence (\ref{ZRnegBpos-equal}). To compute the expectation value 
\be
\widetilde{Z}_{n|p} = \langle 0| ( \overline{W}_p ) ^n |0 \rangle    \label{ZRnegBpos}
\ee
we have to commute all $\G^{t_i}_+$ to the right of all $\G^{t_i}_-$ operators. It is convenient to approach this problem recursively. We therefore assume that manipulating $( \overline{W}_p ) ^n $ to get a form ordered in such a way gives rise to an overall coefficient $C_n$ arising from commutation relations between $\G^{t_i}_{\pm}$. Therefore, in addition moving all $\widehat{Q}$'s to the right:
$$
( \overline{W}_p ) ^n = C_n \Big(\prod_{i=0}^{n-1} \g^1_-(q^i) \Big) \Big(\prod_{i=0}^{n-1} \g^2_+(q^{i-1}) \Big) \widehat{Q}^n.
$$
Now an insertion of one more wall-operator can be written as
\bea
( \overline{W}_p ) ^{n+1} &= & C_{n+1} \Big(\prod_{i=0}^{n} \g^1_-(q^i) \Big) \Big(\prod_{i=0}^{n} \g^2_+(q^{i-1}) \Big) \widehat{Q}^{n+1} = \nonumber \\
& = & \Big[  C_n \Big(\prod_{i=0}^{n-1} \g^1_-(q^i) \Big) \Big(\prod_{i=0}^{n-1} \g^2_+(q^{i-1}) \Big) \widehat{Q}^n \Big] \Big[\g^1_-(x) \g^2_+(x/q) \widehat{Q}\Big] = \nonumber \\
& = & C_n \Big( \prod_{i=0}^{n-1} c_p(q^{n-i}) \Big) \Big(\prod_{i=0}^{n} \g^1_-(q^i) \Big) \Big(\prod_{i=0}^{n} \g^2_+(q^{i-1}) \Big) \widehat{Q}^{n+1}, \nonumber
\eea
where 
$$
c_p(xy) = \prod_{s=1}^p \prod_{r=p+1}^{N+1} \Big(1 - (t_rt_s)\frac{xy}{q_s q_{s+1}\cdots q_{r-1}}  \Big)^{-t_r t_s}
$$ 
arises from a commutation of $\g^2_+(x/q) \g^1_-(y ) = \g^1_-(y ) \g^2_+(x/q) c_p(xy)$. Therefore $\frac{C_{n+1}}{C_n} = \prod_{i=0}^{n-1} c_p(q^{n-i})$ and we get
\bea
\widetilde{Z}_{n|p} & = & C_n = \prod_{k=1}^{n-1} \frac{C_{k+1}}{C_k} = \prod_{k=1}^{n-1} \prod_{i=0}^{k-1} c_p(q^{k-i}) = \prod_{i=1}^{n-1} c_p(q^{n-i})^i = \nonumber \\
& = & \prod_{i=1}^{n-1}  \prod_{s=1}^p \prod_{r=p+1}^{N+1} \Big(1 - (t_rt_s)\frac{q^{n-i}}{q_s q_{s+1}\cdots q_{r-1}}  \Big) ^{-t_r t_s i}.   \nonumber
\eea
which proves (\ref{ZRnegBpos-Results}). A change of variables (\ref{ZRnegBpos-qQ-Results}) leads finally to the identification of crystal and BPS partition functions (\ref{ZRnegBpos-equal}).


\subsubsection*{Chambers with $R>0$, $B>0$}


Here we prove the equality (\ref{ZRposBpos-Results}). We wish to compute
\be
Z_{n|p} = \langle \Omega_+| ( \overline{W}_p ) ^n | \Omega_-\rangle.    \label{ZRposBpos}
\ee
This case is more involved than the previous one with $R<0$, because apart from the ordering of $\G^{t_i}_{\pm}$ in the product of wall-operators, we also have to commute infinite sets of $\G^{t_i}_{\pm}$'s encoded in $|\Omega_{\pm}\rangle$. It is again convenient to approach this problem recursively and determine $Z_{n+1|p}/Z_{n|p}$. First note that we can write
\bea
Z_{n|p} & = & \langle \Omega_+| ( \overline{W}_p ) ^n | \g^1_-(1) \g^2_-(1)  | A_-(q) A_-(q^2)\ldots |0\rangle, \label{Znp-correlator} \\
Z_{n+1|p} & = & \langle \Omega_+| ( \overline{W}_p ) ^n | \g^1_-(1) \g^2_+(q^{-1})  | A_-(q) A_-(q^2)\ldots |0\rangle.  \nonumber
\eea
The only difference between these two expressions is that $\g^2_-(1)$ appears in the former instead of $\g^2_+(q^{-1})$ in the latter. After commuting these operators respectively to the left and to the right we will be left with the same correlator. First of all, the contribution from commuting $\g^2_-(1)$ to the left in $Z_{n|p}$ over $\g^2_+$'s (from all $A_+$ and $\overline{W}_p$) is the same as the contribution from $\g^2_+(q^{-1})$ passing over all $\g^2_-$ in $Z_{n+1|p}$. Then, commuting $\g^2_-(1)$ in $Z_{n|p}$ to the left over all $\g^2_+$ gives a factor
$$
S_n = \prod_{l=0}^{\infty} \prod_{s=1}^p \prod_{r=p+1}^{N+1} \Big(1 - (t_rt_s) q^{l+n+1} q_s q_{s+1}\cdots q_{r-1}  \Big) ^{-t_r t_s },
$$
while commuting $\g^2_+(1)$ to the right in $Z_{n+1|p}$ over all $\g^1_-$ gives
$$
T_{n+1} = \prod_{l=0}^{\infty} \prod_{s=1}^p \prod_{r=p+1}^{N+1} \Big(1 - (t_rt_s) \frac{q^{l+1}}{ q_s q_{s+1}\cdots q_{r-1}}  \Big) ^{-t_r t_s }.
$$

Let us now assume that $Z_{n|p}$ has the form given in (\ref{ZRposBpos-Results})
\be
Z_{n|p} = M(1,q)^{N+1} \, Z^{(0)}_{n|p} \, Z^{(1)}_{n|p} \, Z^{(2)}_{n|p},    \label{Znp-induction}
\ee
with the following factors
\bea
Z^{(0)}_{n|p} & = & \prod_{l=1}^{\infty} \prod_{p \notin \overline{s,r+1} \subset \overline{1,N+1}} \Big(1 - (t_rt_s) \frac{q^{l}}{ q_s q_{s+1}\cdots q_{r-1}}  \Big) ^{-t_r t_s l}  \Big(1 - (t_rt_s) q^{l} q_s q_{s+1}\cdots q_{r-1} \Big) ^{-t_r t_s l} ,    \nonumber \\
Z^{(1)}_{n|p} & = & \prod_{l=1}^{\infty} \prod_{p \in \overline{s,r+1} \subset \overline{1,N+1}}   \Big(1 - (t_rt_s) q^{l+n} q_s q_{s+1}\cdots q_{r-1} \Big) ^{-t_r t_s l} ,    \nonumber \\
Z^{(2)}_{n|p} & = & \prod_{l=n+1}^{\infty} \prod_{p \in \overline{s,r+1} \subset \overline{1,N+1}} \Big(1 - (t_rt_s) \frac{q^{l-n}}{ q_s q_{s+1}\cdots q_{r-1}}  \Big) ^{-t_r t_s l} .     \nonumber 
\eea

Then the relation between correlators in (\ref{Znp-correlator}) implies that
$$
Z_{n+1|p} = M(1,q)^{N+1} \, Z^{(0)}_{n|p} \, Z^{(1)}_{n|p} \, Z^{(2)}_{n|p} \frac{T_{n+1}}{S_n}.
$$
Now we check that
$$
\frac{Z^{(1)}_{n|p} }{S_n} = Z^{(1)}_{n+1|p},\qquad \qquad    Z^{(2)}_{n|p} \, T_{n+1} = Z^{(2)}_{n+1|p}.
$$
The factor $Z^{(0)}_{n|p}=Z^{(0)}_{n+1|p}$ in fact does not depend on $n$, so we conclude that
$$
Z_{n+1|p} = M(1,q)^{N+1} \, Z^{(0)}_{n+1|p} \, Z^{(1)}_{n+1|p} \, Z^{(2)}_{n+1|p}.
$$
This form is consistent with the assumption of the induction proof (\ref{Znp-induction}). This proves (\ref{ZRposBpos-Results}), and further change of variables (\ref{ZRposBpos-qQ-Results}) leads to the identification of crystal and BPS partition functions (\ref{ZRposBpos-qQ-Results}).


\subsubsection*{Chambers with $R<0$, $B<0$}



Here we prove (\ref{ZRnegBneg-Results}). We wish to compute the correlator
\be
\widetilde{Z}'_{n|p} = \langle 0| ( \overline{W}'_p ) ^n | 0\rangle.   \label{ZRnegBneg}
\ee
The proof is analogous as in the case $R<0$ and $B>0$. As we already discussed,
$$
\widetilde{Z}'_{0|p} = \langle 0 | 0 \rangle =  1.
$$
But contrary to the case with $B>0$, an insertion of a single $\overline{W}'_p$ has a non-trivial effect. Again we assume that manipulating $( \overline{W}'_p ) ^n $ to bring all $\G^{t_i}_+$ to the right of $\G^{t_i}_-$ gives rise to an overall coefficient $C'_n$ :
$$
( \overline{W}'_p ) ^n = C'_n \Big(\prod_{i=0}^{n-1} \g^2_-(q^i) \Big) \Big(\prod_{i=0}^{n-1} \g^1_+(q^{-i-1}) \Big) \widehat{Q}^n.
$$
An insertion of one more wall-operator leads to the relation
$$
C'_{n+1} = C'_n \Big( \prod_{i=0}^{n} c'_p(q^{n-i}) \Big) 
$$
where 
$$
c'_p(xy) = \prod_{s=1}^p \prod_{r=p+1}^{N+1} \Big(1 - (t_rt_s) xy q_s q_{s+1}\cdots q_{r-1} \Big)^{-t_r t_s}
$$ 
arises from a commutation of $\g^1_+(x/q) \g^2_-(y ) = \g^2_-(y ) \g^1_+(x/q) c'_p(xy)$. Therefore 
\bea
\widetilde{Z}'_{n|p} & = & C'_n = \widetilde{Z}'_{1|p}  \prod_{k=1}^{n-1} \frac{C'_{k+1}}{C'_k} = \widetilde{Z}'_{1|p} \prod_{k=1}^{n-1} \prod_{i=0}^{k} c'_p(q^{k-i}) = \prod_{i=1}^{n} c'_p(q^{n-i})^i = \nonumber \\
& = & \prod_{i=1}^{n}  \prod_{s=1}^p \prod_{r=p+1}^{N+1} \Big(1 - (t_rt_s) q^{n-i} q_s q_{s+1}\cdots q_{r-1}  \Big) ^{-t_r t_s i}.   \nonumber
\eea
This proves (\ref{ZRnegBneg-Results}). Changing then variables according to (\ref{ZRnegBneg-qQ-Results}) proves (\ref{ZRnegBneg-equal}).


\subsubsection*{Chambers with $R>0$, $B<0$}


In the last case we prove (\ref{ZRposBneg-Results}) and compute
\be
Z'_{n|p} = \langle \Omega_+| ( \overline{W}'_p ) ^n | \Omega_-\rangle.    \label{ZRposBneg}
\ee
The proof is analogous to the case with $B>0$. Again we have to commute all $\G^{t_i}_+$ to the right of $\G^{t_i}_-$.  First note that we can write
\bea
Z'_{n|p} & = & \langle \Omega_+| ( \overline{W}'_p ) ^n | \g^1_-(1) \g^2_-(1)  | A_-(q) A_-(q^2)\ldots |0\rangle, \label{Znpprim-correlator} \\
Z'_{n+1|p} & = & \langle \Omega_+| ( \overline{W}'_p ) ^n | \g^1_+(q^{-1}) \g^2_-(1)  | A_-(q) A_-(q^2)\ldots |0\rangle.  \nonumber
\eea
The only difference between these two expressions is that $\g^1_-(1)$ appears in the former instead of $\g^1_+(q^{-1})$ in the latter. After commuting these operators respectively to the left and to the right we will be left again with the same correlator. The only nontrivial contributions will arise from commuting $\g^1_-(1)$ in $Z'_{n|p}$ to the left over all $\g^2_+$ 
$$
S'_n = \prod_{l=0}^{\infty} \prod_{s=1}^p \prod_{r=p+1}^{N+1} \Big(1 - (t_rt_s) \frac{q^{l+n+1}}{ q_s q_{s+1}\cdots q_{r-1} } \Big) ^{-t_r t_s },
$$
as well as from commuting $\g^1_+(q^{-1})$ to the right in $Z'_{n+1|p}$ over all $\g^2_-$ 
$$
T'_{n+1} = \prod_{l=0}^{\infty} \prod_{s=1}^p \prod_{r=p+1}^{N+1} \Big(1 - (t_rt_s) q^{l} q_s q_{s+1}\cdots q_{r-1}  \Big) ^{-t_r t_s }.
$$

We assume now that $Z'_{n|p}$ has the form given in equation (\ref{ZRposBneg-Results}) 
\be
Z'_{n|p} = M(1,q)^{N+1} \, Z'^{(0)}_{n|p} \, Z'^{(1)}_{n|p} \, Z'^{(2)}_{n|p},    \label{Znpprim-induction}
\ee
with the following factors
\bea
Z'^{(0)}_{n|p} & = & \prod_{l=1}^{\infty} \prod_{p \notin \overline{s,r+1} \subset \overline{1,N+1}} \Big(1 - (t_rt_s) \frac{q^{l}}{ q_s q_{s+1}\cdots q_{r-1}}  \Big) ^{-t_r t_s l}  \Big(1 - (t_rt_s) q^{l} q_s q_{s+1}\cdots q_{r-1} \Big) ^{-t_r t_s l} ,    \nonumber \\
Z'^{(1)}_{n|p} & = & \prod_{l=n}^{\infty} \prod_{p \in \overline{s,r+1} \subset \overline{1,N+1}}   \Big(1 - (t_rt_s) q^{l-n} q_s q_{s+1}\cdots q_{r-1} \Big) ^{-t_r t_s l} ,    \nonumber \\
Z'^{(2)}_{n|p} & = & \prod_{l=1}^{\infty} \prod_{p \in \overline{s,r+1} \subset \overline{1,N+1}} \Big(1 - (t_rt_s) \frac{q^{l+n}}{ q_s q_{s+1}\cdots q_{r-1}}  \Big) ^{-t_r t_s l} .     \nonumber 
\eea

The relation between correlators in (\ref{Znpprim-correlator}) implies that
$$
Z'_{n+1|p} = M(1,q)^{N+1} \, Z'^{(0)}_{n|p} \, Z'^{(1)}_{n|p} \, Z'^{(2)}_{n|p} \frac{T'_{n+1}}{S'_n}.
$$
We now check that
$$
\frac{Z'^{(2)}_{n|p} }{S'_n} = Z^{(2)}_{n+1|p},\qquad \qquad    Z^{(1)}_{n|p} \, T'_{n+1} = Z^{(1)}_{n+1|p}.
$$
The factor $Z'^{(0)}_{n|p}=Z'^{(0)}_{n+1|p}$ does not depend on $n$, and we conclude that
$$
Z'_{n+1|p} = M(1,q)^{N+1} \, Z'^{(0)}_{n+1|p} \, Z'^{(1)}_{n+1|p} \, Z'^{(2)}_{n+1|p}.
$$
This form is consistent with the assumption of the induction proof (\ref{Znpprim-induction}), and therefore (\ref{ZRposBneg-Results}) is proved. Further change of variables (\ref{ZRposBneg-qQ-Results}) leads to the identification of crystal and BPS partition functions (\ref{ZRposBneg-equal}). 




\section{Summary and discussion}   \label{sec-summary}

In this paper we developed a fermionic approach to BPS counting in local, toric Calabi-Yau manifolds without compact four-cycles. We also discussed its crystal melting interpretation, and explained the structure of crystals associated to all manifolds in this class, in a large set of chambers. There are however several issues which require further analysis. 

Firstly, we considered mainly chambers associated to turning on arbitrary $B$-field of magnitude $n$ through one, fixed $p$'th two-cycle in the geometry, in terms of correlators of the wall-crossing operators of the form $\langle (\overline{W}_p)^n \rangle$. In addition we analyzed a few examples of turning on $B$-fields of similar magnitude through all two-cycles simultaneously, and found the corresponding wall-crossing operators $W$ and $V$ in section \ref{ssec-triple}. It would certainly be interesting to find more general wall-crossing operators and associated crystal interpretation for all chambers, i.e. for arbitrary values of $B$-fields through any set of two-cycles turned on simultaneously. 

Secondly, and more conceptually, it would be interesting to find physical reason for the occurrence of these free fermions. For example, such an interpretation has been found in \cite{dhsv} for related, but different fermions arising in the B-model topological vertex \cite{adkmv}. Supposedly our setup would require extension of \cite{dhsv} to include chamber dependence. It is also interesting to see if there is more direct connection between the framework of \cite{adkmv} and ours, and what would be the role of integrable hierarchies in our context.

As we saw in various examples in section \ref{sec-examples}, our crystals unify and generalize various other Calabi-Yau crystal models which appeared previously, such as plane partitions for $\mathbb{C}^3$ and pyramid partitions for the conifold. We also claim that these new crystals are equivalent to crystals introduced in \cite{Ooguri-crystal}, for the class of manifolds that we consider. They must also be related to other statistical models of crystals, dimers, and associated quivers, considered these days in \cite{MozRei,NagaoVO,NagaoYamazaki,Young-dimers,Orbi-vertex}. In particular, modification of crystals upon crossing the walls of marginal stability must translate to the dimer shuffling operation of the corresponding dimer models \cite{Young,MozRei,RefMotQ}. From our results we see that this is obviously true for the conifold. In fact it is not known how to perform dimer shuffling beyond the conifold case. To define it, one could therefore try to translate the effect of the insertion of our wall-crossing operators in the corresponding dimer models. 

We note that there is also a different crystal model for the closed topological vertex \cite{cube}, given in terms of plane partitions in a box whose sides are of finite size and translate to K{\"ahler} parameters of three $\mathbb{P}^1$'s. Furthermore, yet another model for the resolved conifold, related to the deformed boson-fermion correspondence, have been discussed in \cite{deform}. It would be interesting to relate these models to the present results.

There are also further generalizations one might think of. The string coupling $g_s$ could be refined to two independent parameters $\epsilon_1,\epsilon_2$. In certain special cases such refinements were interpreted combinatorially, both for two-dimensional \cite{Nek-Ok} and three dimensional partitions \cite{refined-vertex}. Such a refinement was also found for pyramid partitions for the conifold \cite{RefMotQ} and shown to be equivalent to considering motivic BPS invariants of Kontsevich and Soibelman \cite{KS}. It is tempting to generalize our results in this spirit, both from the perspective of fermionic states as well as crystals. Such generalization could also be translated to refined dimer shuffling operations.

One could also consider open topological string amplitudes \cite{OV99} and their wall-crossing from fermionic and crystal viewpoint. In the large radius chamber open string amplitudes have been interpreted in terms of defects in crystals and analyzed in \cite{va-sa,ps,ps-phd}. It would be nice to extend this interpretation to other chambers and crystals that we propose. The chain of dualities for open string wall-crossing proposed in \cite{cv-wallcross} should also be helpful in finding such an interpretation.
This should also be consistent with the approach to open non-commutative Donaldson-Thomas invariants proposed in \cite{Nagao-open,NagaoYamazaki}.

Finally, one could extend our results to other geometries, such as non-toric manifolds considered in \cite{WallM,HIV,adj}.


\bigskip

\begin{center}
{\bf Acknowledgments}
\end{center}

\medskip

I thank Jim Bryan for discussions which inspired this project. I am also grateful to Robbert Dijkgraaf, Albrecht Klemm, Hirosi Ooguri, Yan Soibelman, Balazs Szendroi, Cumrun Vafa, and Masahito Yamazaki for useful conversations. I appreciate hospitality and inspiring atmosphere of the \emph{Focus Week on New Invariants and Wall Crossing} organized at IPMU in Tokyo, \emph{International Workshop on Mirror Symmetry} organized at the University of Bonn, and \emph{$7^{th}$ Simons Workshop on Mathematics and Physics}, as well as the High Energy Theory Group at Harvard University. This research was supported by the DOE grant DE-FG03-92ER40701FG-02, the Humboldt Fellowship, the Foundation for Polish Science, and the European Commission under the Marie-Curie International Outgoing Fellowship Programme. The contents of this publication reflect only the views of the author and not the views of the European Commission.


\end{document}